\pdfoutput=1

\documentclass[11pt,twoside,a4paper,cmspaper,final,collab]{cms-tdr}
\def\svnVersion{368475}\def\cmsCernNoTag{CERN-EP/2016-045}\def\cmsCernDate{\today}\def\cmsMessage{Published in the Journal of High Energy Physics as \href{http://dx.doi.org/10.1007/JHEP09(2016)027}{\doi{10.1007/JHEP09(2016)027.}}}

\begin{document}\cmsNoteHeader{TOP-13-009}

\hyphenation{had-ron-i-za-tion}
\hyphenation{cal-or-i-me-ter}
\hyphenation{de-vices}
\RCS$Revision: 356157 $
\RCS$HeadURL: svn+ssh://svn.cern.ch/reps/tdr2/papers/TOP-13-009/trunk/TOP-13-009.tex $
\RCS$Id: TOP-13-009.tex 356157 2016-07-15 20:02:31Z alverson $

\newcommand{\mt}{\ensuremath{m_{\ell\nu \PQb}}\xspace}
\newcommand{\cosThetaPol}{\ensuremath{\cos{\theta^*}}\xspace}
\newcommand{\cosThetalbl}{\ensuremath{\cos{\theta_\ell}}\xspace}
\newcommand{\metpt}{\ensuremath{{\textbf{p}\!\!\!/}_{\mathrm{\textbf{\!T}}}}\xspace}
\newcommand{\PFrelIso}{\ensuremath{I_{\mathrm{rel}}}\xspace}
\newcommand{\mTW}{\ensuremath{m_{\mathrm{T}}}\xspace}
\newcommand{\mtw}{\mTW}
\newcommand{\qcd}{multijet\xspace}
\newcommand{\QCD}{\qcd}
\newcommand{\wjets}{\PW+jets\xspace}
\newcommand{\tw}{\PQt\PW\xspace}
\newcommand{\ww}{\PW\PW\xspace}
\newcommand{\wz}{\PW\Z\xspace}
\newcommand{\zz}{\Z\Z\xspace}
\newcommand{\zjets}{\Z+jets\xspace}
\newcommand{\lumivalFBtwelve}{19.7\xspace}
\newcommand{\lumivalFBeleven}{5.1\xspace}
\newcommand{\sigexpcomb}{0.8\xspace}
\newcommand{\sigobscomb}{2.3\xspace}
\newcommand{\sigexpseven}{0.5\xspace}
\newcommand{\sigobsseven}{0.9\xspace}
\newcommand{\sigexpcombfinal}{1.1\xspace}
\newcommand{\sigobscombfinal}{2.5\xspace}
\newcommand{\clsexpmusevenpbnosig}{20.2\xspace}
\newcommand{\clsexpmusevenpbsig}{25.4\xspace}
\newcommand{\clsobsmusevenpb}{31.4\xspace}
\newcommand{\clsexpmusevenpbsiginf}{19.0\xspace}
\newcommand{\clsexpmusevenpbsigsup}{36.6}
\newcommand{\clsexpcombpbnosig}{15.6\xspace}
\newcommand{\clsexpcombpbsig}{20.5\xspace}
\newcommand{\clsobscombpb}{28.8\xspace}
\newcommand{\clsexpcombpbsiginf}{13.4\xspace}
\newcommand{\clsexpcombpbsigsup}{26.7}
\newcommand{\clsexpcombfinalSFnosig}{2.2\xspace}
\newcommand{\clsexpcombfinalSFsig}{3.1\xspace}
\newcommand{\clsobscombfinalSF}{4.7\xspace}
\newcommand{\clsexpcombfinalSFsiginf}{2.1\xspace}
\newcommand{\clsexpcombfinalSFsigsup}{4.0}
\newcommand{\xsecmlemuseven}{7.1\xspace}
\newcommand{\xsecmlemu}{11.7\xspace}
\newcommand{\xsecmleele}{16.8\xspace}
\newcommand{\xsecmlecomb}{13.4\xspace}
\newcommand{\xsecmlecombfinalSF}{2.0\xspace}
\newcommand{\xsecmlecombfinalSFunc}{0.9\xspace}
\newcommand{\xsecmlemusevenerrpb}{8.1\xspace}
\newcommand{\xsecmlemuerrpb}{7.5\xspace}
\newcommand{\xsecmleeleerrpb}{9.1\xspace}
\newcommand{\xsecmlecomberrpb}{7.3\xspace}
\newcommand{\costhetalj}{\cosThetaPol}

\newlength\cmsSkipLength
\setlength\cmsSkipLength{1.5ex}

\cmsNoteHeader{TOP-13-009}
\title{Search for $s$ channel single top quark production in pp collisions at $\sqrt{s} = 7$ and 8\TeV}

\date{\today}

\abstract{
A search is presented for single top quark production in the $s$ channel in proton-proton collisions
with the CMS detector at the CERN LHC in decay modes of the top quark containing a muon or an electron in the final state.
The signal is extracted through a maximum-likelihood fit to the distribution of a multivariate
discriminant defined using boosted decision trees to separate the expected signal contribution from background processes.
The analysis uses data collected at centre-of-mass energies of 7 and 8\TeV
and corresponding to integrated luminosities of \lumivalFBeleven and \lumivalFBtwelve~\fbinv, respectively.
The measured cross sections of $\xsecmlemuseven \pm \xsecmlemusevenerrpb$\unit{pb} (at 7\TeV)
and $\xsecmlecomb \pm \xsecmlecomberrpb$\unit{pb} (at 8\TeV) result in a best fit value
of $\xsecmlecombfinalSF \pm \xsecmlecombfinalSFunc$ for the combined ratio of the measured 
and expected values.
The signal significance is \sigobscombfinal standard deviations, and the
upper limit on the rate relative to the standard model expectation is \clsobscombfinalSF at 95\% confidence level.
}

\hypersetup{
pdfauthor={CMS Collaboration},
pdftitle={Search for s channel single top quark production in pp collisions at sqrt(s) = 7 and 8 TeV},
pdfsubject={CMS},
pdfkeywords={CMS, physics, top quarks}}

\maketitle

\section{Introduction}
\label{sec:introduction}
\label{sec:Introduction}
\begin{figure}[htb]
       	  \centering
	    \includegraphics[width=0.20\textwidth]{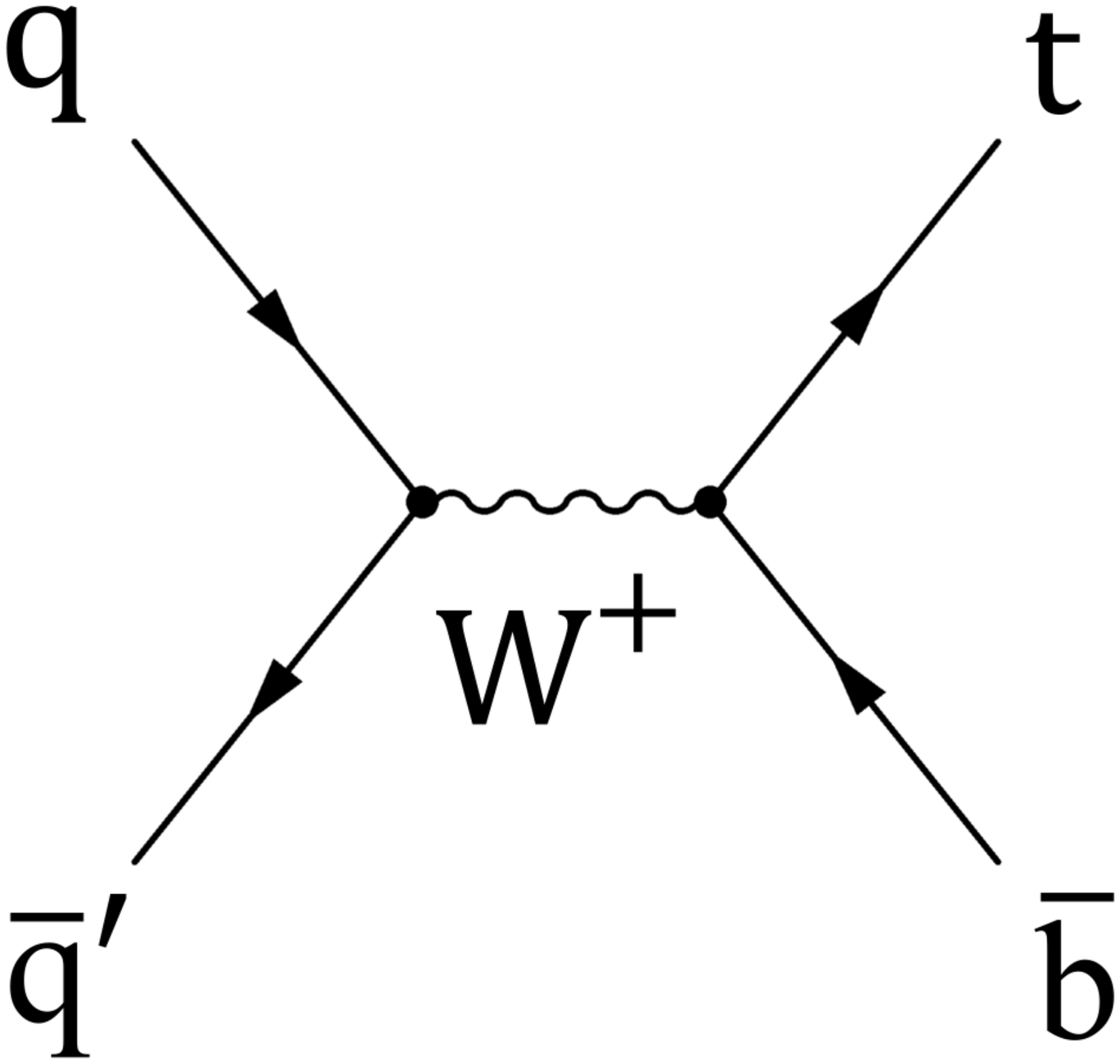}\hfil
	    \includegraphics[width=0.20\textwidth]{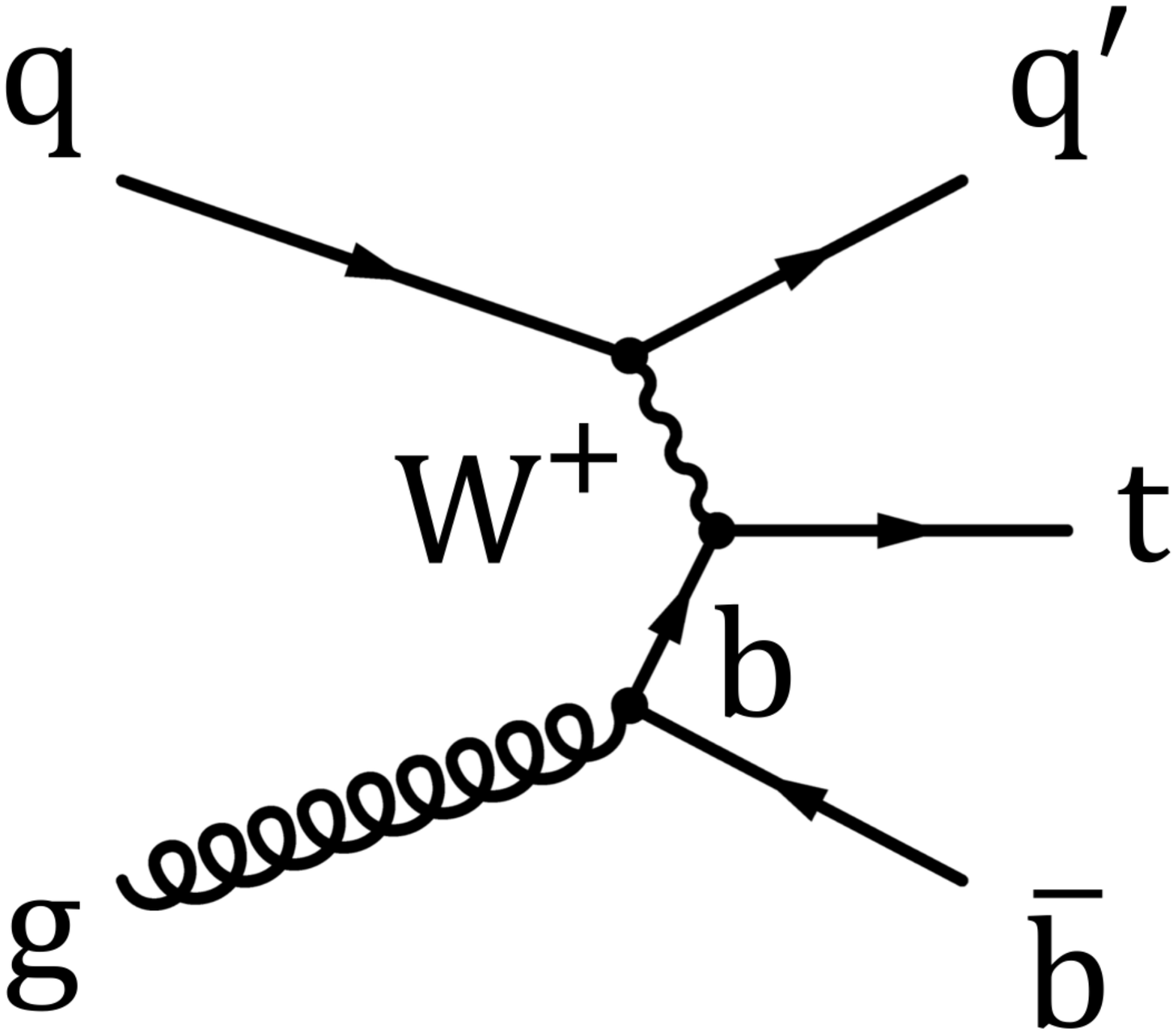}\hfil
	    \includegraphics[width=0.20\textwidth]{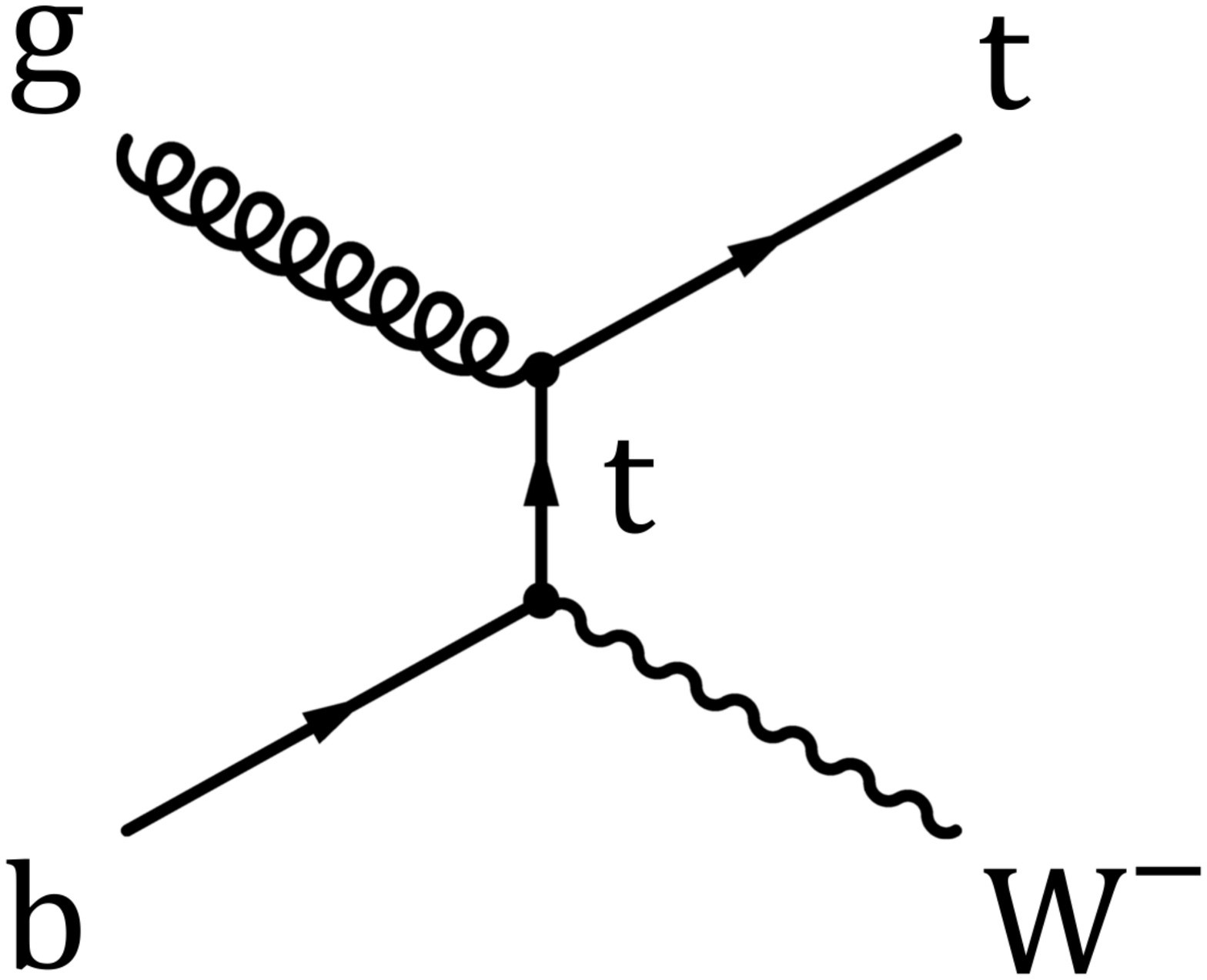}
	    \caption{\label{fig:FG}Leading-order Feynman diagram for single top quark production in (left) the $s$
	    channel, whose production rate is studied in this paper, (middle) the dominant
            next-to-leading-order diagram in the $t$ channel, and (right) the tW production channel.}
	\end{figure}

Top quarks at the CERN LHC are produced mainly in pairs through the strong interaction, but
can also be produced individually via a charged-current electroweak interaction. The study of single top quark production thereby provides
probes of the electroweak sector of the standard model (SM), which predicts three production channels:
the $s$ channel, the $t$ channel, and the W-associated or tW production channel (Fig.~\ref{fig:FG}).

The first observations of single top quark production were announced
by the D0 and CDF collaborations at the Fermilab Tevatron in 2009~\cite{PhysRevLett.103.092001,Aaltonen:2009jj}.
Evidence for $s$ channel production was announced by the D0 collaboration in 2013~\cite{D0paper},
while the process was definitively observed when combining the searches from both the D0
and the CDF collaborations~\cite{PhysRevLett.112.231803}.
Evidence for $s$ channel production was confirmed by the ATLAS Collaboration at the LHC~\cite{s-channel-Atlas-8TeV-evidence},
where the search is challenging because the process is suppressed in proton-proton (pp) collisions.

For pp collisions at $\sqrt{s} = 7$ and 8\TeV, the SM predicted $s$ channel cross sections are
\begin{equation*}\begin{aligned}
\label{eq:crosssection_tot}
\sigma_{s} (7\TeV) &= 4.56\pm0.07\,\text{(scale)}\pm 0.17\,\mathrm{(PDF)}\unit{pb}, \text{ and}\\
\sigma_{s} (8\TeV) &= 5.55\pm0.08\,\text{(scale)}\pm 0.21\,\mathrm{(PDF)}\unit{pb},
\end{aligned}\end{equation*}	
as calculated in quantum chromodynamics (QCD) at approximate next-to-next-to-leading order (NNLO), including
resummation of soft-gluon emission within next-to-next-to-leading logarithms (NNLL)~\cite{Kidonakis:2012db}.
The first uncertainty corresponds to a doubling and halving of the renormalization and
factorization scales. The second uncertainty is from the choice of parton distribution functions (PDFs)
at the 90\% confidence level (CL).

All three single top quark production channels, shown in Fig.~\ref{fig:FG}, are directly related to the Cabibbo-Kobayashi-Maskawa
matrix element $V_{\PQt\PQb}$, providing a direct measurement of this SM parameter.
The $s$ channel production process is of special interest since a possible deviation from the
SM prediction of its cross section may indicate the presence of mechanisms beyond the standard model (BSM),
as predicted by models that involve the exchange of a non-SM mediator, such as a $\PWpr$
boson or a charged Higgs boson~\cite{Hashemi:2013raa}.
A review of deviations from SM predictions for
$s$ and $t$ channel modes in BSM scenarios
can be found in Ref.~\cite{Tait:2000sh}.

This paper presents a search performed at the CMS experiment for single top quark production
in the $s$ channel considering the leptonic decay channels of the W boson produced
in top quark decay. Only the decays of the W boson into a muon or an electron ($\ell=\mu$, e) and a
corresponding neutrino are considered.
Decays of the W boson into a tau lepton and a neutrino, where the tau lepton subsequently decays into a muon or an electron,
are regarded as part of the signal.
Events are selected considering the kinematic properties of physical objects reconstructed in the
final state. Three statistically independent analysis categories are therefore defined, according to
the number and flavour of the reconstructed jets.
Dedicated strategies are used in data to estimate and reject \QCD backgrounds.
The procedure for signal extraction consists of a
simultaneous fit to the distributions of multivariate discriminants
trained separately in each analysis category on a set of kinematic variables
that show separation between signal and background.

This measurement is performed using LHC pp collision data collected by the CMS detector
corresponding to the integrated luminosities of \lumivalFBeleven and \lumivalFBtwelve~\fbinv
at centre-of-mass energies of 7 and 8\TeV, respectively.
While at 7\TeV only the muon channel is considered, at 8\TeV both the muon and electron channels are included.

\section{The CMS detector}
\label{sec:detector}
The central feature of the CMS apparatus is a
superconducting solenoid of $6$\unit{m} internal
diameter providing an axial magnetic field of 3.8\unit{T}.
The inner region accommodates the silicon pixel and strip tracker
which records charged particle trajectories with high granularity
and precision up to pseudorapidity $\abs{\eta}=2.5$.
An electromagnetic calorimeter (ECAL) made of lead tungstate crystals
and a brass and scintillator sampling hadron calorimeter,
both arranged in a barrel assembly and two endcaps, surround
the tracking volume and extend up to the region $\abs{\eta}<3.0$.
Coverage up to $\abs{\eta}=5.0$ is provided by a quartz-fibre and steel
absorber Cherenkov calorimeter.
Muons are measured in gas-ionization detectors embedded in the steel
flux-return yoke outside the solenoid.
A more detailed description of the CMS detector, together with a definition
of the coordinate system and the relevant kinematic variables,
can be found in Ref.~\cite{JINST}.

\section{Simulated samples}
\label{sec:samples}

The nominal $s$ channel single top quark events in this study are generated using the
next-to-leading order (NLO) \POWHEG~1.0~\cite{Frixione:2007vw} event generator.
The CTEQ6.6M program~\cite{CTEQ6} is used to model the proton PDF. The top quark mass
is set to 172.5\GeV, and tau lepton decays are modelled with TAUOLA~\cite{TAUOLA}.
For the 7\TeV analysis, a large sample of signal events
generated using the leading-order (LO) matrix-element \COMPHEP~4.4~\cite{comphep}
generator is employed for the training of the multivariate discriminant.
The generators are interfaced to LO \PYTHIA~6.4 (Z2 tune)~\cite{Sjostrand:2006za} for showering and hadronization.
Monte Carlo (MC) simulated events with a single top quark are normalized to the approximate NNLO+NNLL cross
section of 3.14\unit{pb}at 7\TeV and 3.79\unit{pb}at 8\TeV~\cite{Kidonakis:2012db}.
MC simulated events with a top antiquark are normalized to the approximate NNLO+NNLL cross section of
1.42\unit{pb}at 7\TeV and 1.76\unit{pb}at 8\TeV.
The other single top quark processes, $t$ channel, and
tW production, are considered as backgrounds for this measurement and
are simulated using the \POWHEG~1.0 generator.

The main background in this analysis is top quark pair production (\ttbar)
in final states with one or two charged leptons.
Single vector bosons in association with jets, \wjets, and \zjets, are also included in the background.
Both \ttbar\ and single vector boson events are generated using LO matrix element
\MADGRAPH~5.1~\cite{madgraph5new} interfaced to \PYTHIA~6.4.
The background from diboson (WW, ZZ, and WZ) events is small and is generated with \PYTHIA~6.4.
Multijet background events from QCD processes 
are extracted directly from data or from a simulated sample generated with \PYTHIA~6.4 (see Section~\ref{sec:BDTqcd}).
The cross sections for the background processes in the analysis are summarized in Table~\ref{tab:samples}.

The cross sections are reported at approximate NNLO+NNLL accuracy for single top quark~\cite{Kidonakis:2012db}
and $\ttbar$ production~\cite{Czakon:2013}, at NNLO accuracy for $\Z/\gamma^*$+jets
and $\PW$+$n$ jets (with $n = 1, 2, 3,$ and $4$) events~\cite{PhysRevD.86.094034},
and at the LO level for the remaining contributions. When stated, the cross section includes
the branching ratio of the leptonic decay, including electrons, muons, and tau leptons.
The multijet sample is defined by the presence of at least one generator-level muon with $\pt > 15\GeV$,
and requiring the transverse momentum generated in the hard scattering parton process to be greater than 20\GeV.

         \begin{table}
         \centering
         \topcaption{Monte Carlo cross sections calculated for background processes.}
         \begin{tabular}{ lrrr }
         \hline
         Process        & \multicolumn{1}{c}{$\sigma$\,[pb] at 7\TeV} & \multicolumn{1}{c}{$\sigma$\,[pb] at 8\TeV}  \\
         \hline
         Single top quark ($t$ channel)          & 43.0   & 56.4  \\
         Single antitop quark ($t$ channel)      & 22.9   & 30.7  \\
         Single top or antitop quark ($\tw$)     & 7.8    & 11.1 \\[\cmsSkipLength]
        \ttbar                                                & 172.0          & 245.8 \\
        $\PW(\to \ell \nu)+$1 jet                            & 4500           & 5400 \\
        $\PW(\to \ell \nu)+$2 jets                           & 1400           & 1800  \\
        $\PW(\to \ell \nu)+$3 jets                           & 300            & 520 \\
        $\PW(\to \ell \nu)+$4 jets                           & 170            & 210 \\
        $\Z/\gamma^*(\to \ell^+\ell^-)+$jets                & 3000           & 3500 \\
        $\ww$                                                    & 43             & 57 \\
        $\wz$                                                    & 18             & 32 \\
        $\zz$                                                    & 5.9            & 8.3 \\
        $\mu$-enriched multijet events                           & $85\,000$      &  --- \\
        \hline
        \end{tabular}
        \label{tab:samples}

   \end{table}

For all generated processes, the detector response is simulated using a
detailed description of the CMS detector, based on \GEANTfour~\cite{geant}.
A reweighting procedure is applied to simulated
events to reproduce the distribution of the number of multiple pp
interactions per bunch crossing (pileup events) observed in data.

\section{Selection and reconstruction}
\label{sec:selection}
\label{sec:Selection}
The final-state topology in the $s$ channel is characterized by the presence of
one isolated muon or electron, a neutrino that results in an imbalance in the transverse momentum 
of the event, and two b quarks, one originating from the top quark decay
and one recoiling against the top quark. 

Events with at least one muon were selected by the online trigger~\cite{JINST}, requiring 
$\pt>17\GeV$ at 7\TeV, $\pt>24\GeV$ at 8\TeV, $\abs{\eta}<2.1$, and lepton isolation criteria.
Similarly, for electrons at 8\TeV, the corresponding values are $\pt>27\GeV$ and $\abs{\eta}<2.5$.

Because of the increase in instantaneous luminosity during the second part of the 7\TeV run, 
the single muon trigger had to be prescaled and was replaced by a hadronic trigger that required 
at least one muon as defined above and at least one jet in the central region of the detector 
with $\pt>30\GeV$, satisfying an online b tagging criterion.
Simulated leptonic trigger efficiencies are corrected to match those measured in data.
Hadronic trigger efficiencies are not simulated but are measured in data and parametrized as a 
function of the jet $\pt$ in order to reweight the simulated events.

At least one primary vertex is required to be reconstructed from at least four tracks
and to satisfy $\abs{z_\mathrm{PV}}<24$\unit{cm} and $\rho_{\rm PV}<2$\unit{cm},
where $\abs{z_\mathrm{PV}}$ and $\rho_{\rm PV}$ are the respective longitudinal and transverse distances
of the primary vertex relative to the center of the detector.
When more than one interaction vertex is found,
the one with largest sum in $\pt^{2}$ of associated tracks is defined as the primary vertex.

The particle candidates are required to originate from the primary vertex,
and are reconstructed using the CMS particle-flow (PF) algorithm~\cite{CMS-PAS-PFT-09-001}.
Reconstructed muons with $\pt >20\GeV$ at 7\TeV and $\pt >26$\GeV at 8\TeV
within the trigger acceptance ($\abs{\eta} < 2.1$) are selected for analysis.
At 8\TeV, reconstructed electrons~\cite{Khachatryan:2015hwa} with
$\pt >30$\GeV within $\abs{\eta} < 2.5$ are selected, excluding the transition region
between ECAL barrel and endcaps ($1.44 < \abs{\eta} < 1.57$)
where the reconstruction of electrons is not optimal.

Lepton isolation is applied using the $\PFrelIso$ variable, defined as the ratio
between the sum of the transverse energies (\ET) of stable charged
hadrons, photons, and neutral hadrons
in a cone of size $\Delta R = \sqrt{\smash[b]{(\Delta\eta)^2+(\Delta\phi)^2}}$
around the lepton direction (where $\phi$ is the azimuth in radians),
and the $\pt$ of the lepton.
At 7\TeV, the muon isolation requirement is $\PFrelIso < 0.15$ with $\Delta R = 0.3$.
At 8\TeV, $\PFrelIso$ is corrected by subtracting the average contribution from neutral
particles in pileup events. It is required $\PFrelIso < 0.12$ with $\Delta R = 0.4$ for muon isolation,
and $\PFrelIso < 0.1$ with $\Delta R = 0.3$ for electron isolation.

The presence of a single muon or electron satisfying the criteria described above is required to
reduce the contribution from dilepton events,
which can arise from \ttbar or from
$\qqbar \to \ell^+ \ell^-$+jets Drell--Yan (DY) processes.
Events containing additional muons or electrons, with looser requirements for muons of $\pt >10\GeV$
within the full acceptance of $\abs{\eta} < 2.5$, and $\PFrelIso < 0.2$, and for electrons with
$\pt > 20$\GeV, $\abs{\eta} < 2.5$, and $\PFrelIso < 0.15$ are rejected.

Jets are reconstructed using the anti-\kt algorithm~\cite{Cacciari:2008gp} with a distance parameter of 0.5,
using as input the particles identified through the PF algorithm.
To reduce contamination from pileup events, charged particle candidates not associated with the primary vertex
are excluded from the jet reconstruction. The energies of jets are corrected by the estimated amount
of energy deposited in the jet area~\cite{jetarea-2008} from pileup hadrons. Scale factors depending on the \ET and $\eta$
of the jets~\cite{CMS-PAS-JME-10-003} are further applied and reflect the detector response.
The analysis considers jets within $\abs{\eta}<4.5$ and $\pt>40$\GeV.
We identify jets stemming from b quarks through b tagging algorithms~\cite{csvbtag}.
The threshold on the discriminant value is set to provide a misidentification probability (mistag)
for light-parton jets of about 0.1\%. The corresponding
b tagging efficiency ranges from 40 to 60\%, depending
on jet $\pt$ and $\eta$ and on the specific algorithm.
Simulated b tagging efficiencies are corrected to match those measured in data~\cite{csvbtag, CMS-PAS-BTV-13-001}.

The imbalance in transverse momentum (vector $\metpt$) is defined as the projection on the plane
perpendicular to the beams of the negative of the vector sum of the momenta of all
reconstructed particles in an event.
Its magnitude is referred to as $\ETslash$.
It is assumed that the $x$ and $y$ components of the missing momentum, $(\metpt)_x$ and $(\metpt)_y$,
are entirely due to the escaping neutrino.
The longitudinal component $p_{z,\nu}$ of the neutrino momentum is estimated from a quadratic equation obtained by imposing
that the invariant mass of the lepton-neutrino system must be equal to the invariant mass of the W boson.
In case of two real solutions, the smallest $p_{z,\nu}$ is chosen, while when two complex solutions are found
the imaginary part is eliminated by recalculating $(\metpt)_x$ and $(\metpt)_y$ independently,
to provide a W boson with a transverse mass of 80.4\GeV.
The W boson transverse mass is defined as
\begin{equation*}
\mTW = \sqrt{\left(p_{\mathrm{T},\ell} + p_{\mathrm{T},\nu}\right)^2 - \left( p_{x,\ell} + p_{x,\nu} \right)^2 - \left(  p_{y,\ell} + p_{y,\nu} \right)^2},
\end{equation*}
where $p_{\mathrm{T},\ell}$ and $p_{\mathrm{T},\nu}$ are the lepton and neutrino transverse momenta and
$p_{x,\ell}$, $p_{y,\ell}$, $p_{x,\nu}$ and $p_{y,\nu}$ are the components of the
lepton and neutrino transverse momenta along the $x$ and $y$ axes.
Finally, four-momenta of top quark candidates are reconstructed from the lepton 
and the jet originating from the b quark produced in top quark decay, using also the quantities $\metpt$ and $p_{z,\nu}$.
In events with more than 1 b jet, the one which results in a reconstructed top mass closer to the nominal 
one is chosen.

The selected events are classified into statistically independent ``$N$-jets $M$-tags'' analysis categories,
where $N$ refers to the number of reconstructed jets above 40\GeV and $M$ to the number of selected jets
passing the b tagging requirement.
Three event categories are used for this analysis:
the 2-jets 2-tags category is $s$ channel enriched, and employed in signal extraction,
the 2-jets 1-tag category is useful to constrain the $t$ channel and \wjets backgrounds,
while the 3-jets 2-tags category is useful to constrain the dominant \ttbar\ background.
In each event category, further requirements are applied to reject the multijet background, which 
in the 8 TeV analysis is separated from the other components by means of a QCD BDT discriminator. 
The strategies to reject the multijet background and to estimate its contribution 
will be described in Section~\ref{sec:BDTqcd}.

An additional selection is applied in the 8\TeV signal 2-jets 2-tags category
that exploits the property of $s$ channel events to have a
lower number of additional jets with $20 < \pt < 40\GeV$ (loose jets)
than \ttbar\ events. Only events with no more than 1 loose jet are selected.
The requirement selects 60\% of \ttbar events and 90\% of $s$ channel events.

Because of the presence of two b-tagged jets in the final state,
the 2-jets 2-tags and the 3-jets 2-tags categories are reconstructed with
a top quark candidate for each of the two b jets.
The candidate with invariant mass closest to the nominal
top quark mass of 172.5\GeV is then selected for further study in the analysis.
Using this method, the efficiency of association of the correct
b jet to the top quark is measured to be 74\%
in $s$ channel events and 70\% in \ttbar events.
The dependence of the correct b jet association on top quark mass
is evaluated in $s$ channel events by changing the top quark
mass by the conservative estimation of its uncertainty of $\pm$1.5\GeV,
which yields changes in efficiency of less than 1\%.

\section{Implementation of the multivariate analysis}
\label{sec:BDT}
Since the SM prediction for the signal yield is much smaller than the background processes,
it is important to enhance the separation between signal and background events
to measure the $s$ channel with highest possible significance.
A multivariate analysis was therefore developed, in
which boosted decision tree (BDT) discriminants~\cite{ref:bdt} are defined
for each event category, based on a set of input discriminants.
In this section the BDTs for signal extraction are described, while in the next section the BDTs
for the multijet background rejection will be presented.

The BDT training and the choice of the input discriminants is performed separately 
for the muon channels at 7 and 8\TeV and for the electron channel at 8\TeV,
taking into account the different selections and the different level of background,
in particular for the \QCD background.
The samples employed for training and evaluation of performance are taken from simulation,
with the exception of the \QCD background, which is
taken from a data control sample, as described in Section~\ref{sec:BDTqcd}.

Several discriminants are investigated
for possible input to the BDTs, in particular kinematic and angular variables exploiting the
properties of $s$ channel events~\cite{PhysRevD.81.034005}.
For each channel, the set of input variables are defined according to the following
criteria. A variable must be well modelled in simulation, and must significantly increase the discrimination
power of a BDT (after comparing performance of the BDTs trained without it).

The most important variables chosen as input to the BDTs in the
2-jets 2-tags category are: \mtw, the angular separation between the two jets
($\Delta R_{\PQb\PQb}$), the invariant mass of the system composed of the lepton and
subleading jet ($m_{\ell\PQb2}$), the transverse momentum of the two-jet system
($\pt^{\mathrm{bb}}$), and the difference in azimuthal angle between the top quark and the leading jet ($\Delta \phi_{\PQt,\PQb1}$).
The leading and subleading jets refer to the two jets with largest $\pt$.

The other variables used as input to the BDTs are
the invariant mass of the top quark candidate in the event (\mt),
the scalar sum of the \pt of all jets ($H_\mathrm{T}$),
the cosine of the angle between lepton and the beam axis in the top quark rest frame (\cosThetalbl),
\ETslash, the lepton \pt, and the difference in azimuthal angle between the top quark and the next-to-leading b jet
($\Delta \phi_{\PQt,\PQb2}$).

\begin{figure}[!htb]
\centering
\includegraphics[width=0.45\textwidth]{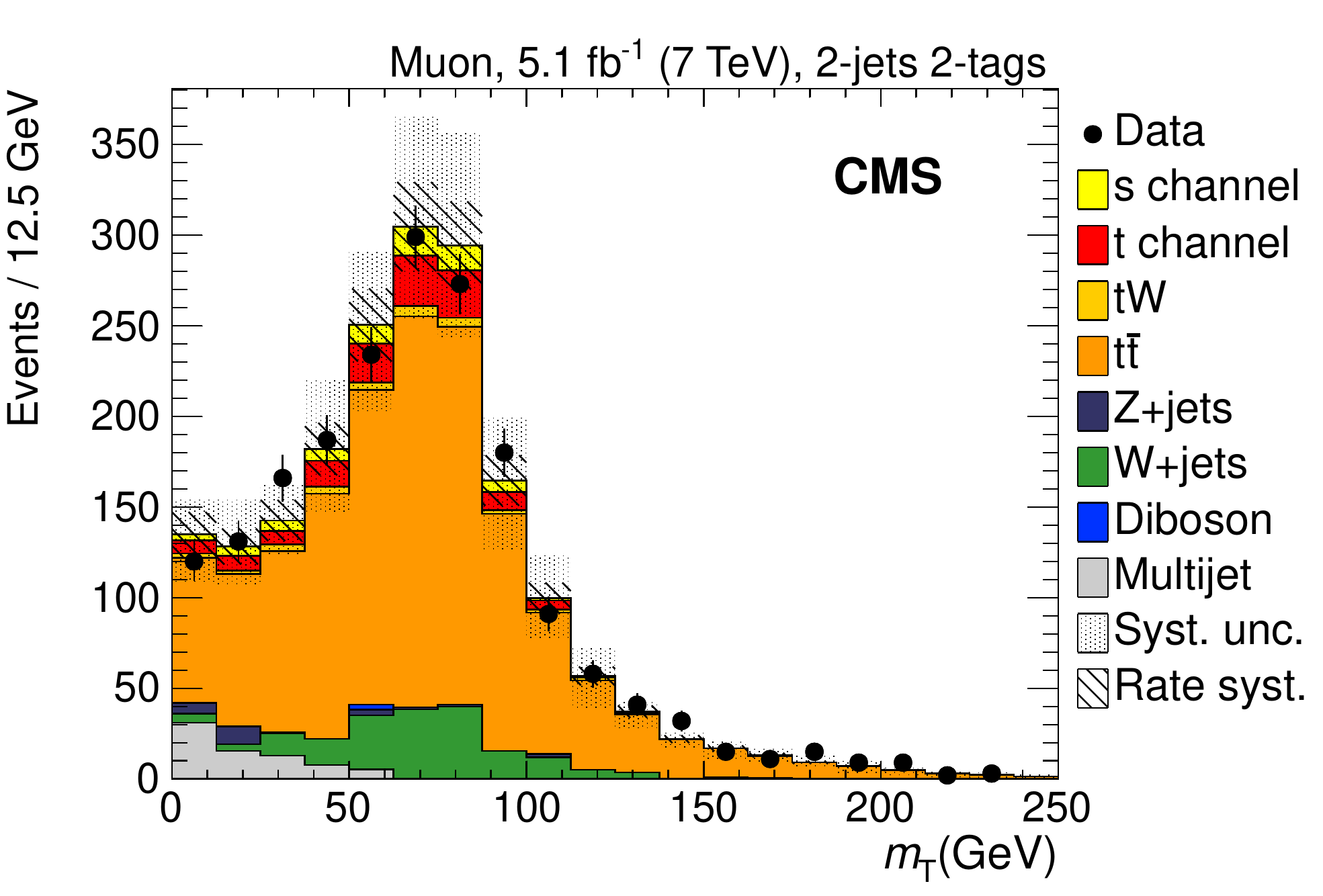}
\includegraphics[width=0.45\textwidth]{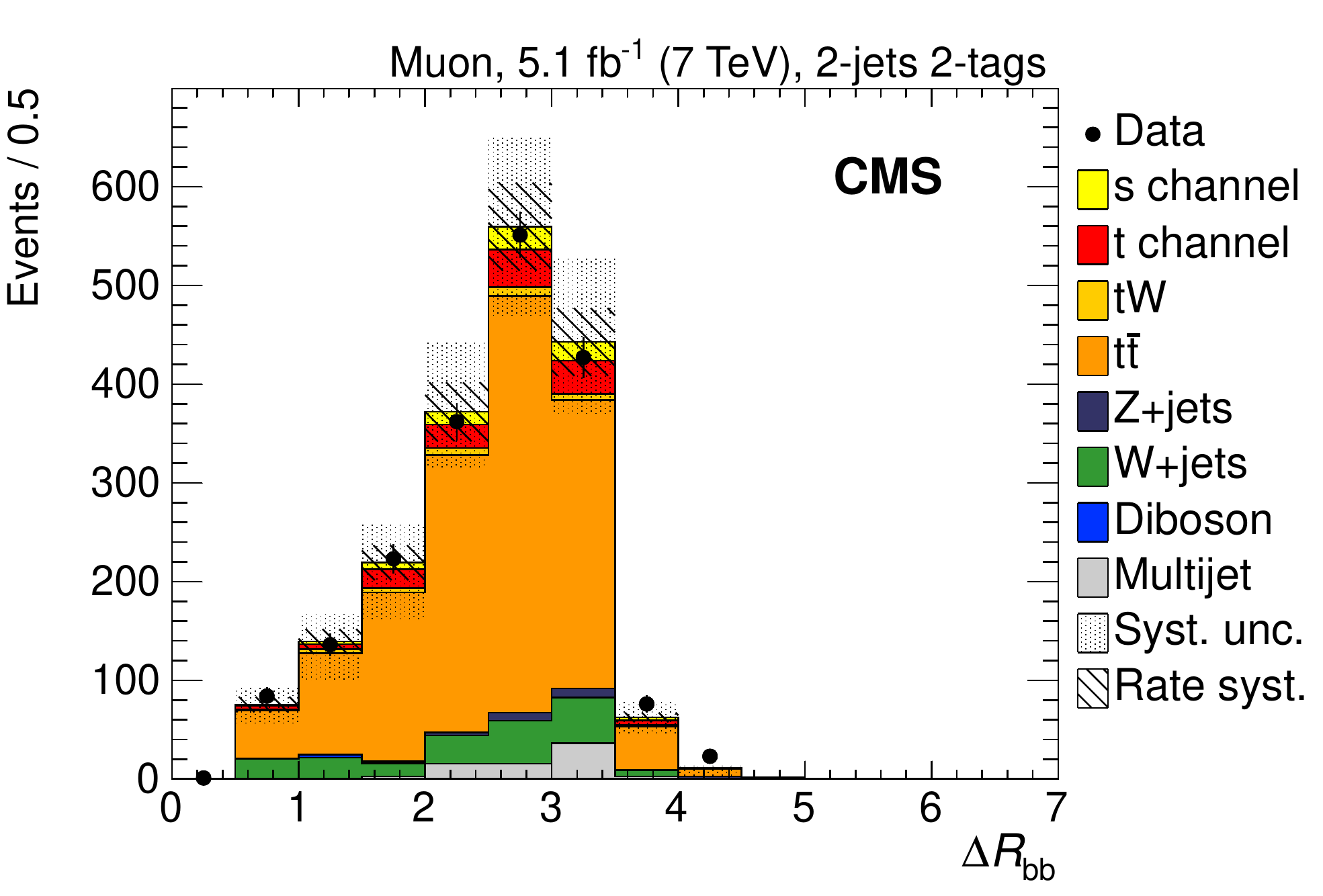}
\includegraphics[width=0.45\textwidth]{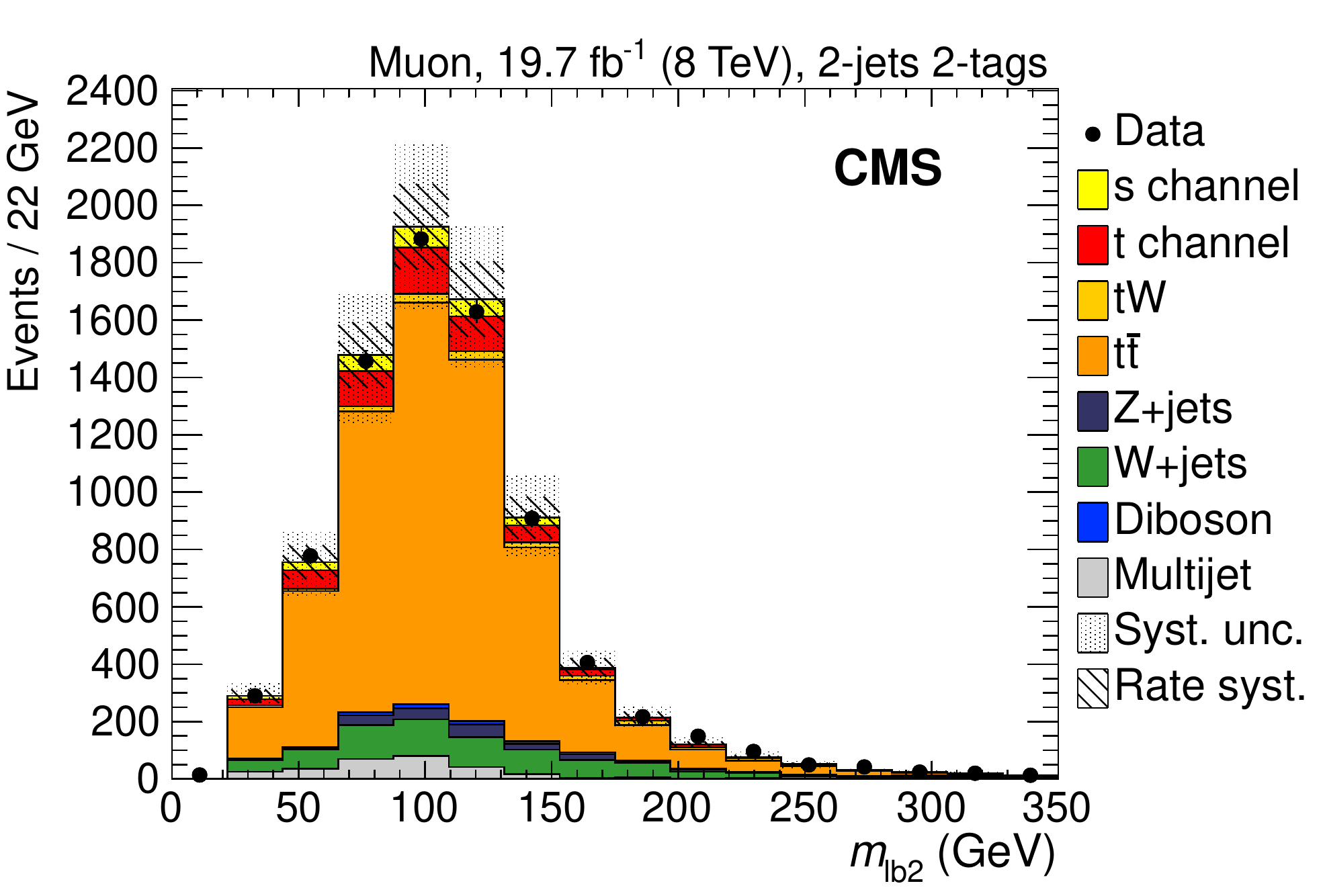}
	      \includegraphics[width=0.45\textwidth]{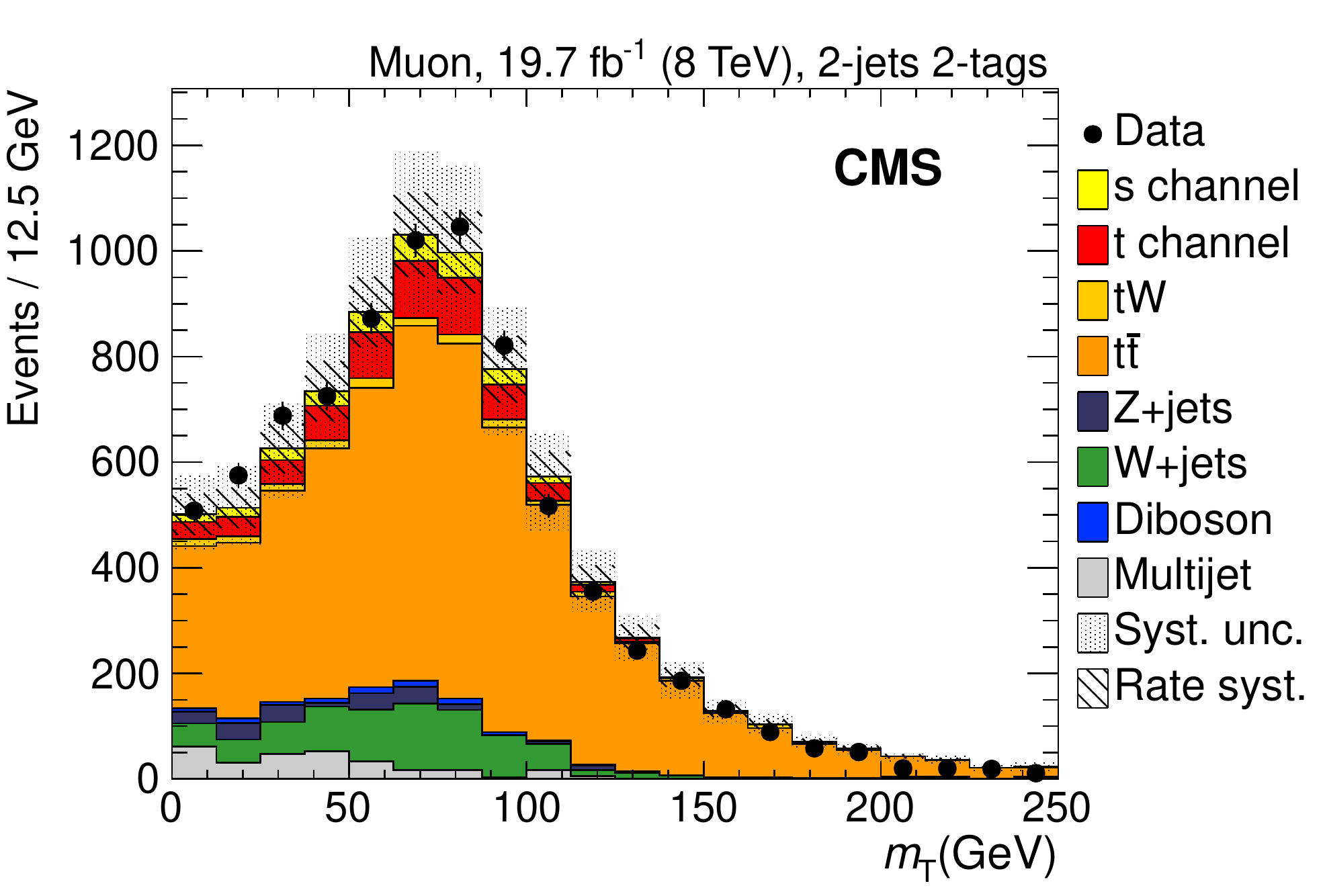}
	      \includegraphics[width=0.45\textwidth]{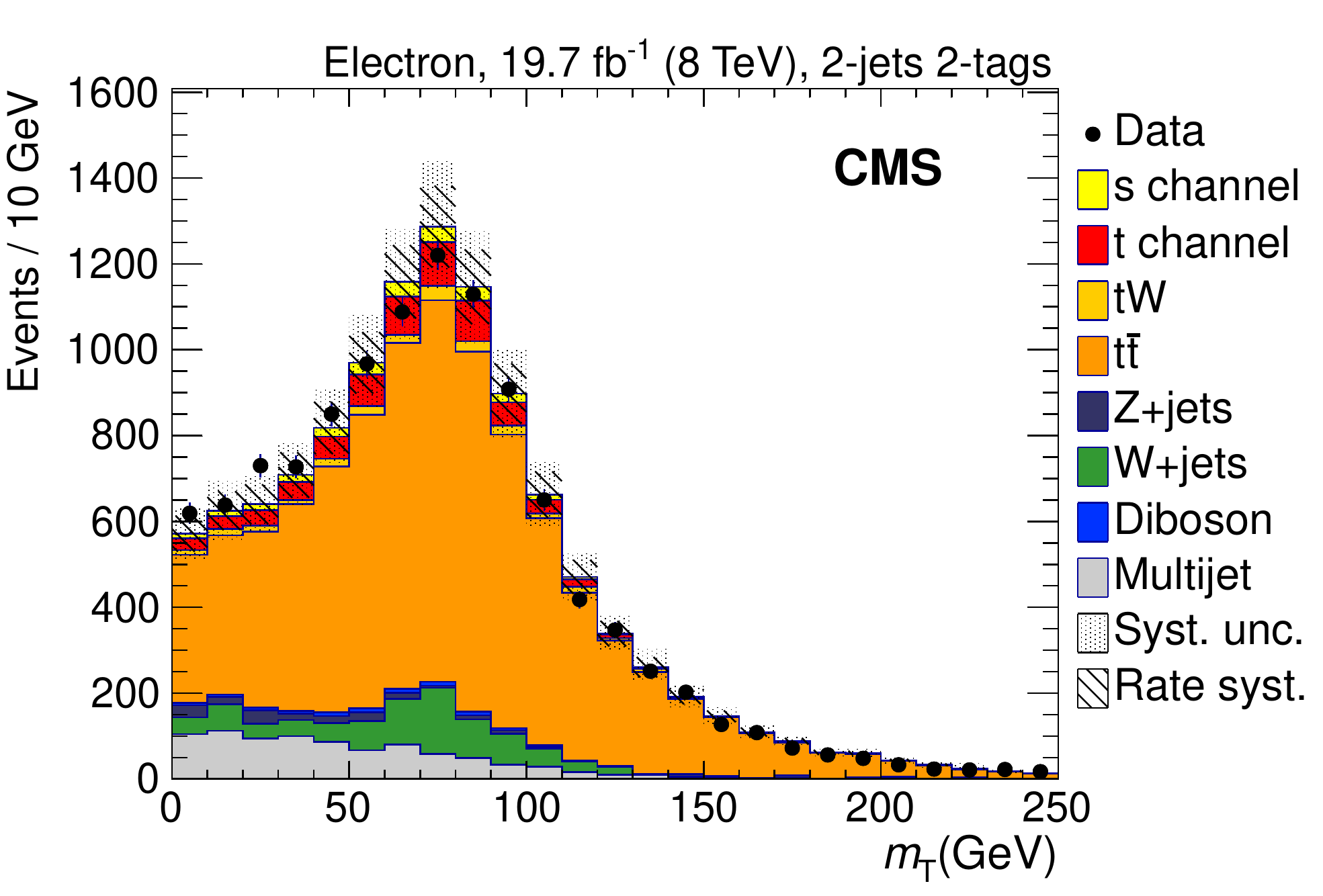}
	      \includegraphics[width=0.45\textwidth]{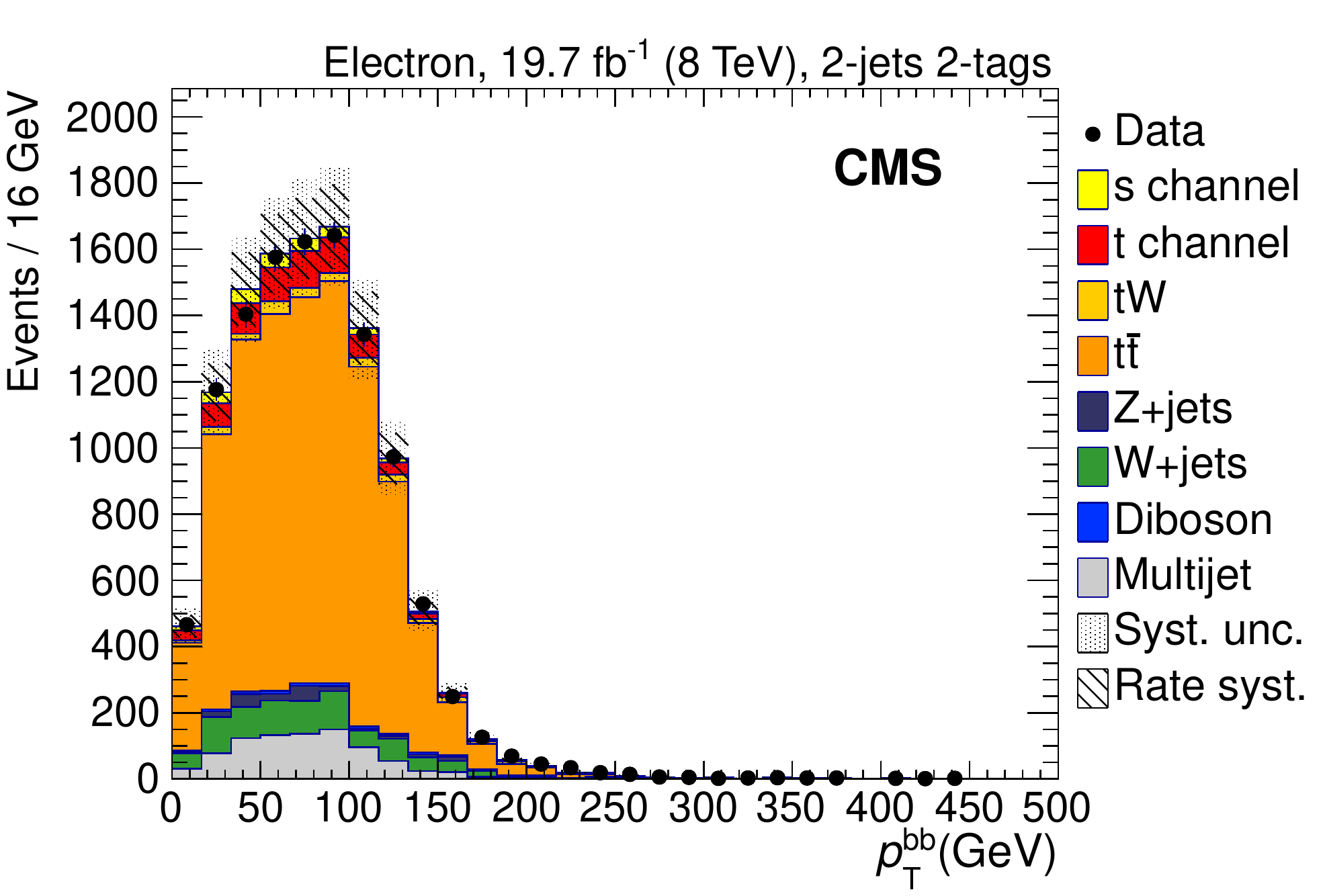}
	    \caption{\label{fig:inputvars-3channels} Comparison between data and simulation in distributions
                           of highest-ranked variables
                           in the 2-jets 2-tags category: (upper left) \mtw and (upper right) $\Delta R_{\PQb\PQb}$
                           for the muon channel at 7\TeV,
	                   (middle left) $m_{\ell \mathrm{b2}}$ and (middle right) \mtw
                           for the muon channel at 8\TeV,
                           and (bottom left) \mtw and (bottom right) $\pt^{\PQb\PQb}$
                           for the electron channel at 8\TeV.
                           The simulation is normalized to the data and the \QCD background is normalized
                           through the maximum-likelihood fit discussed in Section~\ref{sec:BDTqcd},
                           prior to rejecting the \QCD background.
                           The smaller error bands represent only the systematic uncertainties on the 
                           background normalizations, while the larger ones include the total systematic
                           uncertainty obtained from the sum in quadrature of
                           the individual contributions listed in
                           Section~\ref{sec:syst}.}
\end{figure}

Figure~\ref{fig:inputvars-3channels} shows the comparison between data and MC events for the highest ranked variables,
where the simulation is normalized to the number of events selected in data.

The most important variables chosen as input to the BDTs in the
2-jets 1-tags category are: 
the angular separation between the two jets ($\Delta R_{\PQb\PQq}$), 
the cosine of the angle between the lepton and the jet recoiling against the top quark in the 
top quark rest frame ($\cos\theta^\ast$),
\mt,
the invariant mass of the two-jet system ($m_{\PQb\PQq}$),
and $H_\mathrm{T}$.
The other variables are the invariant mass of the system composed of the lepton and
subleading jet ($m_{\ell,\mathrm{j}2}$), 
the lepton pseudorapidity ($\eta_\ell$),
and the difference in azimuthal angle between the $\metpt$ and the lepton ($\Delta \phi_{\metpt,\ell}$).

The most important variables chosen as input to the BDTs in the
3-jets 2-tags category are:
$\pt^{\PQb\PQb}$,
$m_{\ell\PQb2}$, 
the cosine of the angle between the lepton and the non b-tagged jet in the top quark rest frame ($\cos\theta^\ast_{\PQq}$),
and \mtw.
The other variables are
\mt,
$H_\mathrm{T}$,
the transverse momentum of the next-to-leading b jet ($\pt^{\PQb2}$),
and the transverse momentum of the non b-tagged jet ($\pt^{\PQq}$).

\section{Multijet background}
\label{sec:BDTqcd}
In the 7\TeV analysis, the W boson $\mTW$ distribution is employed
to discriminate against the \QCD background.
Multijet events populate the lower part of the $\mTW$ spectrum and the
requirement $\mTW>50$\GeV is applied to suppress their contribution to a negligible
level in the 2-jets 1-tag event category. The number of \QCD events that pass the selection is
estimated from simulation. In the other categories,
the level of \QCD production is already small compared to other backgrounds,
and its contribution is estimated through a maximum-likelihood fit to
the $\mTW$ distribution.

In the 8\TeV analysis, BDT discriminants, referred to as QCD BDTs, are used to reject \QCD events
following the same procedure as in Section~\ref{sec:BDT}.
For each event category a QCD BDT is trained using \QCD events as signal against non-\QCD processes,
and the distribution of the QCD BDT discriminant in data is employed to define a \QCD-enriched interval.
Events with the discriminant value in this interval are rejected from the analysis. The number of rejected
\QCD events is estimated through a maximum-likelihood fit to the QCD BDT distribution
in the \QCD-enriched interval in data.
This number, multiplied by a scale factor obtained from the selection acceptance, provides
the yield of remaining \QCD events for each category.

The most important variables chosen as input to the QCD BDTs in the
2-jets 2-tags category are: lepton \pt, lepton $\eta$, \mt, \mtw,
\cosThetaPol, 
and the transverse momentum of the leading b jet ($\pt^{\PQb}$).
The distributions for the multijet background are extracted from a data sample enriched with
such events.
In the muon channel, the sample is defined by
an anti-isolation requirement on the muon ($0.2 < \PFrelIso < 0.5$ at 7\TeV and
$\PFrelIso > 0.2$ at 8\TeV). In the electron channel, it is defined by requiring
the failure either of the isolation criteria or the tight identification criteria
on the electron.
Since the number of events in the \QCD-enriched data sample at 7\TeV
is lower than at 8\TeV due to smaller integrated luminosity,
no QCD BDT is defined in the 7\TeV analysis.

Table~\ref{tab:BDTqcdcuts} presents the $s$ channel and \QCD event acceptances of
the QCD BDT selection. Different acceptances are observed in the different event
categories since the QCD BDT selection is optimized to minimize the loss of signal events.

\begin{table}[!htb]
\centering
\topcaption{QCD BDT selection acceptance for multijet and $s$ channel events at 8\TeV.\label{tab:BDTqcdcuts}}
   \begin{tabular}{ clcc }
     \hline
     \multirow{2}{*}{Lepton} & \multirow{2}{*}{Event category}  & \multicolumn{2}{c}{Acceptance (\%)} \\ \cline{3-4}
                    &     &   Multijet   & $s$ channel \\
     \hline
     \multirow{3}{*}{$\mu$} &  2-jets 1-tag  & 38 &  75 \\
                            &  2-jets 2-tags & 50 &  92 \\
                            &  3-jets 2-tags & 30 &  74 \\[\cmsSkipLength]
  \multirow{3}{*}{e}     &  2-jets 1-tag  & 29 &  58 \\
                            &  2-jets 2-tags & 60 &  92 \\
                            &  3-jets 2-tags & 40 &  68 \\
\hline
\end{tabular}
\end{table}

\begin{figure}[htb]
	  \centering
            \includegraphics[width=0.48\textwidth]{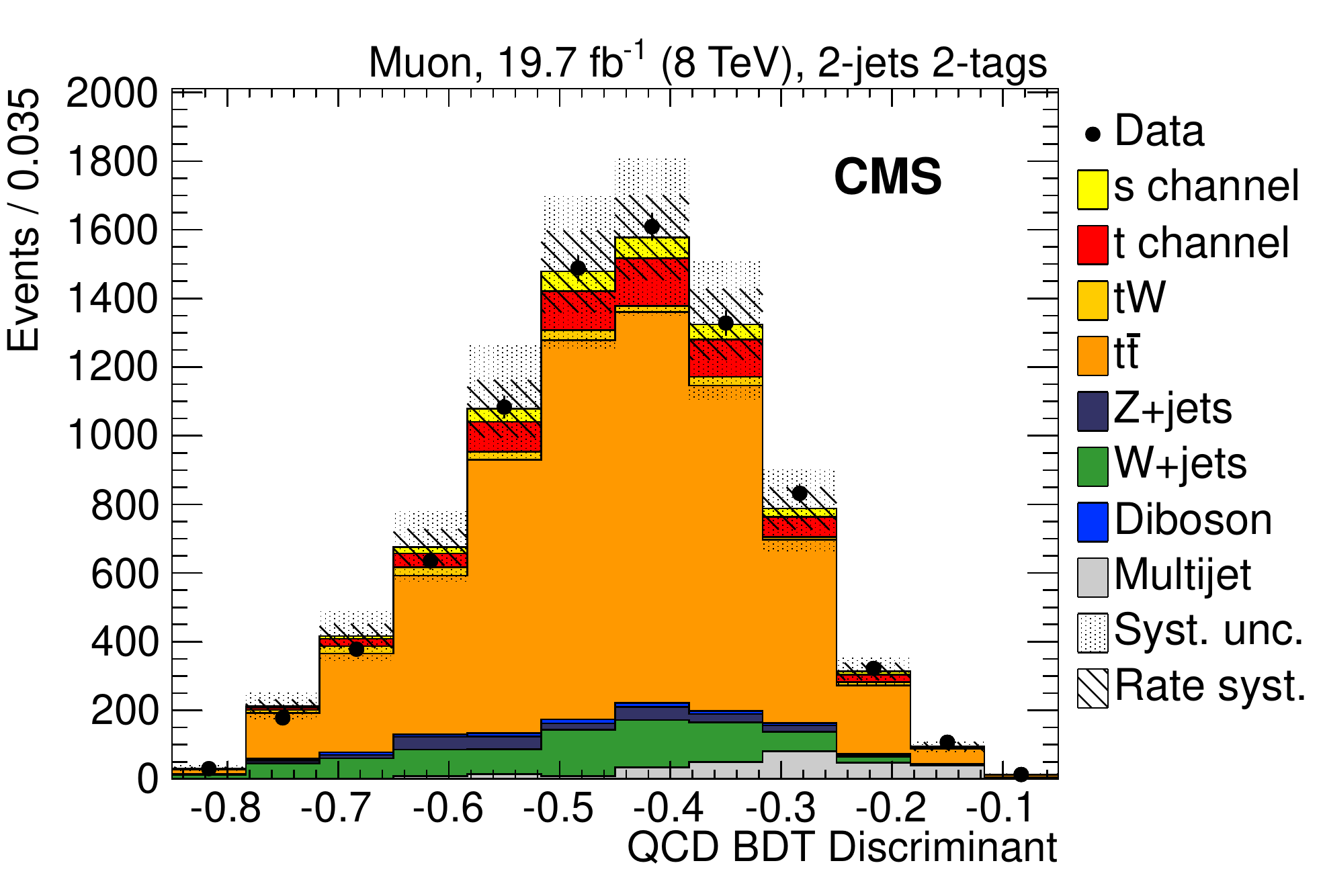}
            \includegraphics[width=0.48\textwidth]{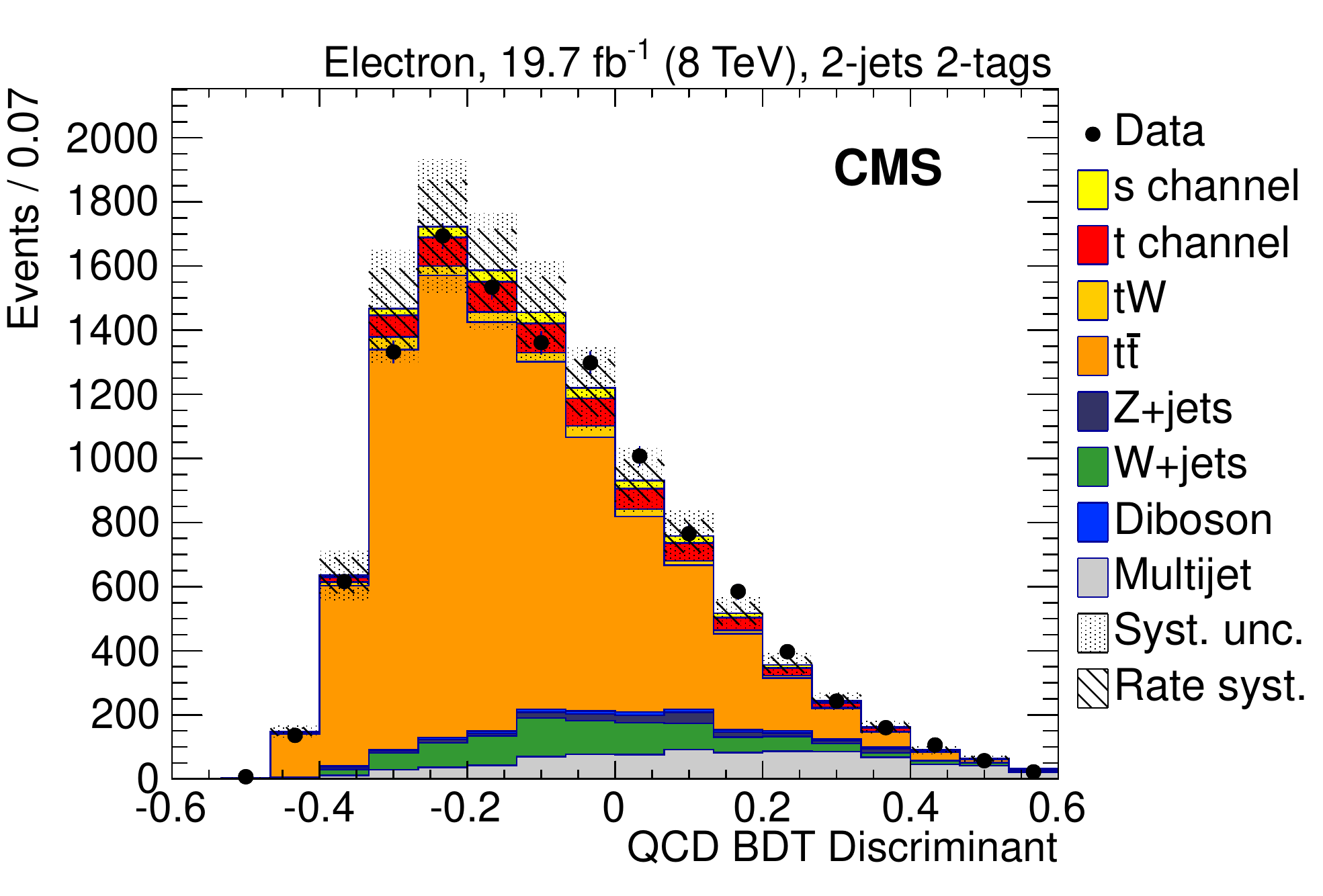}
	    \caption{\label{fig:QCDbdt}
              Comparison of data with simulation for distributions in the QCD BDT discriminant
              in the 2-jets 2-tags event category, in (left) the muon and (right) electron channel at 8\TeV.
              The simulation is normalized to the data.
              While the smaller error bands include the systematic uncertainties on the 
              background normalizations only, the larger ones include the total systematic uncertainty
              obtained summing in quadrature the individual contributions
              discussed in Section~\ref{sec:syst}.}
\end{figure}

Figure~\ref{fig:QCDbdt} shows a comparison of the distributions
in QCD BDT discriminants in data and simulation in the 2-jets 2-tags category
for muon and electron channels at 8\TeV, where the simulation
is normalized to events in data.

Both in 7 and 8\TeV analyses (except for 2-jets 1-tag category at 7\TeV)
a maximum-likelihood fit is performed to determine the yield in \QCD events.
We define the parametrized function $F(x)= a\, V(x)+b\, M(x)$, where $x$ represents
the discriminant variable
and $V(x)$ and $M(x)$ are the respective distributions (templates) in the sum of all processes including a W or Z boson in the
final state, or multijet events. The $V(x)$ distribution is taken from simulation, while $M(x)$ is the template based on the \QCD-enriched data sample.

The total uncertainty on the \QCD background is obtained by
considering the statistical uncertainty
from the fit and possible systematic contributions,
which are evaluated by repeating the fit
after changing the sum of non-\QCD components
by $20\%$ and employing a \QCD template model taken from an independent
sample in data, where neither of the two jets pass the b tagging requirement.

\section{Systematic uncertainties}
\label{sec:syst}
\label{sec:systs}
\label{sec:systematics}
Several sources of systematic uncertainties have been investigated
and determined as follows.
Uncertainties on the normalization are summarized in Table~\ref{tab:rateunc}. Uncertainties on \ttbar~and \wjets are based
on the CMS measurements~\cite{JHEP022014} and~\cite{wjets2014204}, respectively. We refer to a 7\TeV measurement of relative
uncertainty in \wjets, since it represents the most recent result within CMS of the W boson production cross section in association
with two b jets. Uncertainties on \zjets and dibosons come from Refs.~\cite{JHEP06120} and~\cite{mcfm}, respectively, while the uncertainties
on single top quark tW production and $t$ channel are taken from Refs.~\cite{Kidonakis:2012db,JHEP06090,PhysRevLett.112.231802}.
Uncertainties on the \QCD background normalization reported in the table 
come from the extraction procedure described in Section~\ref{sec:BDTqcd}.

\begin{table}[!tbh]
\centering
\topcaption{Summary of normalization uncertainties on the background processes. 
The uncertainties on the \QCD background refer to the 2-jets 2-tags, 2-jets 1-tag, and 3-jets 2-tags categories, respectively.
\label{tab:rateunc}}
\begin{tabular}{ lc }
\hline
 Process                            & Uncertainty (\%) \\
\hline
 \ttbar                             &   10     \\
 \wjets                             &   20     \\
 \zjets                             &   20     \\
 Diboson                            &   30     \\
 tW                                 &   15     \\
 $t$ channel                        &   10     \\
 Multijet, $\mu$, 7\TeV           &   30, 100, 100     \\
 Multijet, $\mu$, 8\TeV           &   30, 10, 30       \\
 Multijet, e, 8\TeV                &   20, 5, 25        \\
\hline
\end{tabular}
\end{table}

The uncertainties on jet energy scale (JES) and jet energy resolution (JER) are taken into account in line with Ref.~\cite{Chatrchyan:2011ds}.
The ``unclustered energy'' in the event, which is computed by subtracting from the $\metpt$ the negative vector sum of the
uncorrected transverse momenta of jets and leptons not clustered in jets, is changed by 10\%.
For each of these changes the $\ETslash$ is recalculated accordingly.
The uncertainties in lepton-reconstruction and trigger-efficiency scale factors are measured using DY events.
The parametrizations describing the hadronic trigger efficiencies are varied and new weights
are applied to simulated events in order to estimate the hadronic trigger uncertainty.
The scale factors used to correct simulation to reproduce the b
tagging efficiency and the mistag fraction measured in data are
changed by their measured uncertainties~\cite{CMS-PAS-BTV-13-001}.

The uncertainty in the total number of interactions per bunch crossing (5\%) is propagated to the modelling of pileup
in the simulated samples. The integrated luminosity is known to an uncertainty of 2.2\% for the 7\TeV data~\cite{lumi2012}
and 2.6\% for the 8\TeV data~\cite{lumi2013}.

The uncertainty from the choice of factorization and
renormalization scales $\mu_{\mathrm{F}}$ and $\mu_{\mathrm{R}}$ in the QCD calculation
is based on dedicated simulated samples of
\ttbar, single top quark production in $s$ channel and $t$ channel, and \wjets events,
with $\mu_{\mathrm{F}}$ and $\mu_{\mathrm{R}}$ varied from half to twice their nominal values.
The uncertainty from matching matrix element and parton shower
thresholds is determined from simulated samples
of \ttbar\ and \wjets with parton matching
threshold doubled and halved relative to their nominal values.
The uncertainty on the chosen set of PDF is estimated by reweighting the simulated events with each
of the 52 eigenvectors of the CT10 PDF parametrization~\cite{PDF:CTEQ10}.

Differential measurements have shown that the \pt spectrum
of the top quarks in \ttbar events is significantly softer than the one generated using MC simulation programs~\cite{toppt}.
Scale factors for event reweighting are derived from these measurements.
The $s$ channel cross section is remeasured based on
samples without any reweighting and samples
that have been reweighted with doubled weights,
as an indication of the corresponding uncertainty.
The effect of the limited number of events in the simulated samples has been
taken into account using the ``Barlow-Beeston light'' method~\cite{barlow_beeston}.

\section{Cross section extraction}
\label{sec:xsection}
A binned maximum-likelihood fit is performed to the BDT data distributions in the
2-jets 2-tags, 2-jets 1-tag, and 3-jets 2-tags categories simultaneously.
In particular, the inclusion in the fit of the 2-jets 1-tag and 3-jets 2-tags regions largely constrains
the \wjets and the \ttbar\ backgrounds respectively while taking into account all possible correlations in the systematic
uncertainties for the three samples.
The expected total yield $\lambda_i$ in each bin $i$ of the BDT distribution is given by the sum of all
the background contributions $B_{p,i}$ and the signal yields $S_{i}$ scaled by the signal-strength modifier
$\beta_{\text{signal}}$, which is defined as the ratio
between the measured signal cross section and the SM prediction, as
\begin{equation*}
 \lambda_{i}(\beta_{\text{signal}}, \theta_u) = \beta_{\text{signal}} \, S_{i} + \sum_{p} c_{p}(\theta_u) B_{p,i}.
\label{eq:model_syst}
\end{equation*}
The modelling of BDT distributions for the $s$ channel and for
each background process $p$, $S$, and $B_p$,
are scaled to the integrated luminosity of the data according to the
SM cross sections.
The uncertainty in each background normalization, except for \QCD events,
is included in the likelihood model through a ``nuisance'' parameter with a
log-normal prior ($c_{p}(\theta_u)$).
The \QCD component is instead fixed to the value estimated with the
method described in Section~\ref{sec:BDTqcd}.

The measured $s$ channel cross section is given by the value of $\beta_{\text{signal}}$ at which
the logarithm of the likelihood function  reaches its maximum. The 68\% CL interval for
the cross section is evaluated by profiling the logarithm of the likelihood as a function of
$\beta_{\text{signal}}$, and taking the parameter values for which the profile likelihood is
0.5 units below its maximum.

The impact from the systematic uncertainty in the background normalizations on the $s$ channel cross section is evaluated
by removing one nuisance at a time from the likelihood model and measuring the corresponding change
in the total uncertainty.
The impact of the uncertainties that are not included in the fit
are evaluated using the following procedure.
For each systematic effect two pseudo-experiments are generated
by changing the corresponding quantity by $+1$ and $-1$ standard deviation.
Maximum-likelihood fits are then performed for each of the pseudo-experiments,
and the differences between the fitted $\beta_{\text{signal}}$ and the nominal one
are taken as the corresponding uncertainties.

The uncertainties arising from different systematic sources are combined according to Ref.~\cite{barlow_article}.
A breakdown of contributions to the overall uncertainty in the measurement is reported in Table~\ref{tab:systtot}.

\begin{table}[!htb]
\centering
\topcaption{Summary of the relative impact of the statistical and systematic uncertainties on the cross section measurement. Different prior
uncertainties have been assigned to \ttbar, single top quark $t$ channel and tW production, \wjets, \zjets and diboson normalizations, see
Section~\ref{sec:systematics}\label{tab:systtot}.}
\begin{tabular}{ lrrrrr }
\hline
 Source   & \multicolumn{5}{c}{Uncertainty (\%)} \\ \cline{2-6}
            &                            \multicolumn{1}{c}{$\mu$, 7\TeV} & \multicolumn{1}{c}{$\mu$, 8\TeV} & \multicolumn{1}{c}{\Pe, 8\TeV}  & \multicolumn{1}{c}{$\mu+\Pe$, 8\TeV}  & \multicolumn{1}{c}{7$+$8\TeV} \\
\hline
Statistical &                             34  &  15 &  14  &  10  &  11 \\[\cmsSkipLength]

 \ttbar, single top quark normalization & 29 &  15 &  14 &  12 &  14  \\
 W/Z+jets, diboson normalization &        23 &  11 &  13 &  12 &  12 \\
 Multijet normalization &                  9  &  3  &  5  &  2  &  2  \\[\cmsSkipLength]

 Lepton efficiency &                      14 &  1  &  2  &  1  &  3  \\
 Hadronic trigger &                       5  &  ---  &  ---  &  ---  &  1  \\
 Luminosity &                             10 &  5  &  6  &  4  &  6  \\
 JER \& JES &                             66 &  39 &  29 &  34 &  18 \\
 b tagging \& mistag &                    34 &  15 &  14 &  14 &  16 \\
 Pileup &                                 6  &  11 &  7  &  9  &  7  \\
 Unclustered \ETslash &                   5  &  8  &  2  &  6  &  5  \\
 $\mu_\mathrm{R} , \mu_\mathrm{F}$ scales &   54 &  34 &  31 &  30 &  28 \\
 Matching thresholds &                    43 &  11 &  12 &  7  &  17 \\
 PDF &                                    12 &  8  &  7  &  7  &  9  \\
 Top quark \pt reweighting &              3  &  5  &  7  &  6  &  6  \\[\cmsSkipLength]
Total uncertainty  &                      115 &  64 &  54 &  55 &  47 \\
\hline
\end{tabular}
\end{table}

\begin{figure}[!htb]
         \centering
           \includegraphics[width=0.48\textwidth]{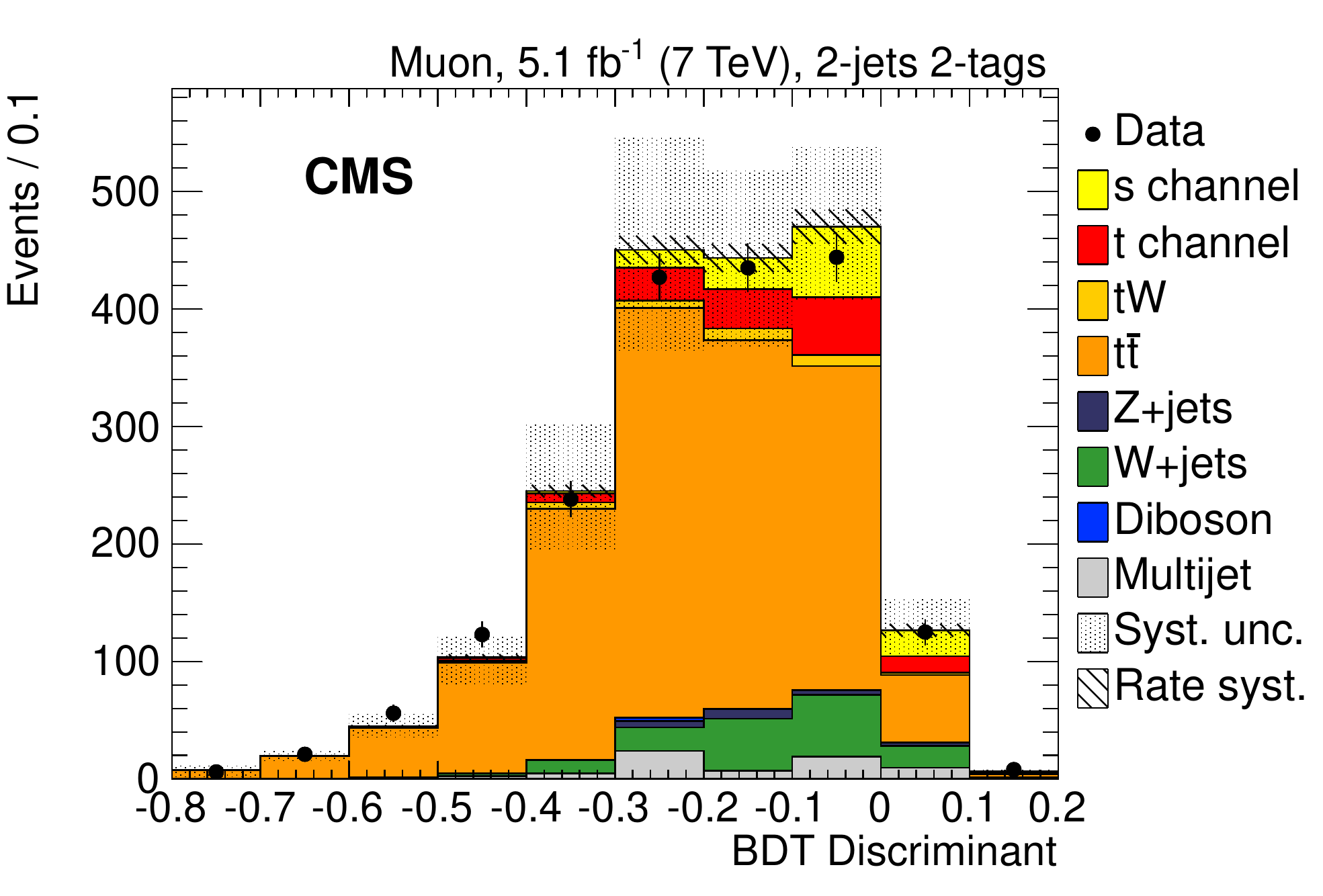}
           \includegraphics[width=0.48\textwidth]{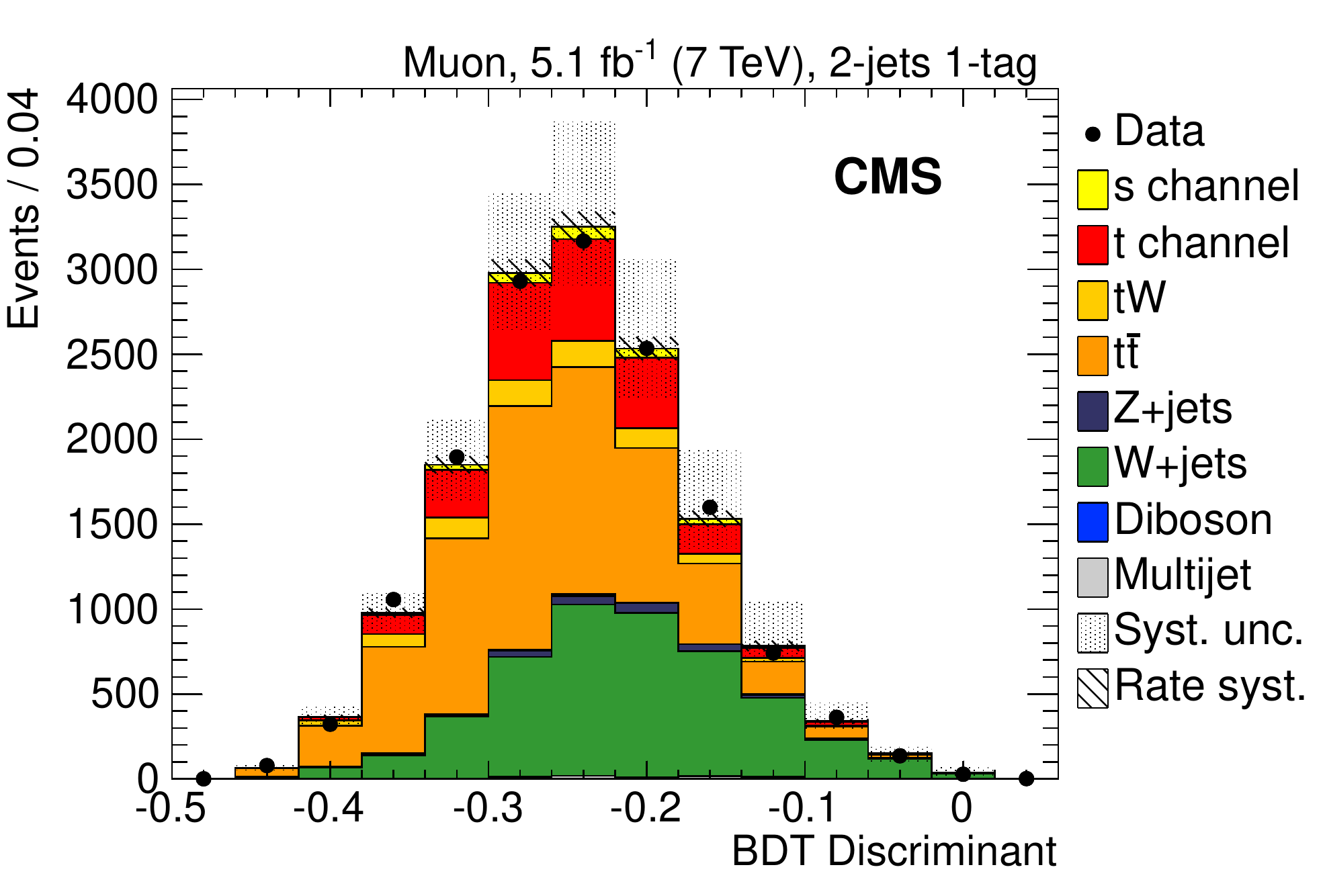}
           \includegraphics[width=0.48\textwidth]{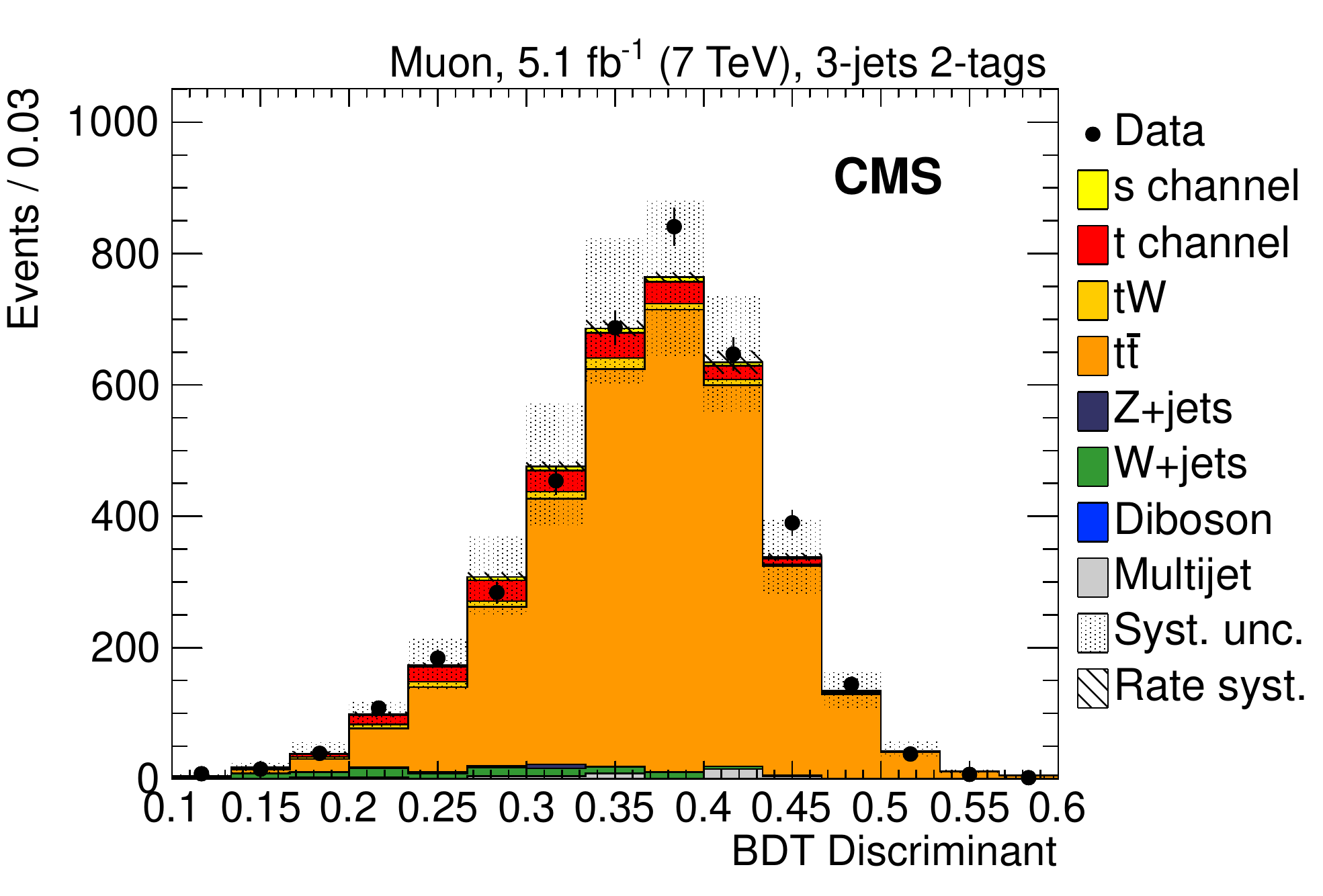}
            \caption{\label{fig:bdtdataMCMufit7TeV}Comparison of data with simulation for distributions of the
              BDT discriminants in the (upper left) 2-jets 2-tags, (upper right) 2-jets 1-tag, and (bottom) 3-jets 2-tags event category,
              for the muon channel at 7\TeV. The simulation is normalized to the combined (7+8\TeV) fit results.
	      The inner uncertainty bands include the post-fit background normalizations uncertainties only, the outer ones
              include the total systematic uncertainty obtained summing in quadrature the individual contributions.}
       \end{figure}

\begin{figure}[!htb]
         \centering
           \includegraphics[width=0.48\textwidth]{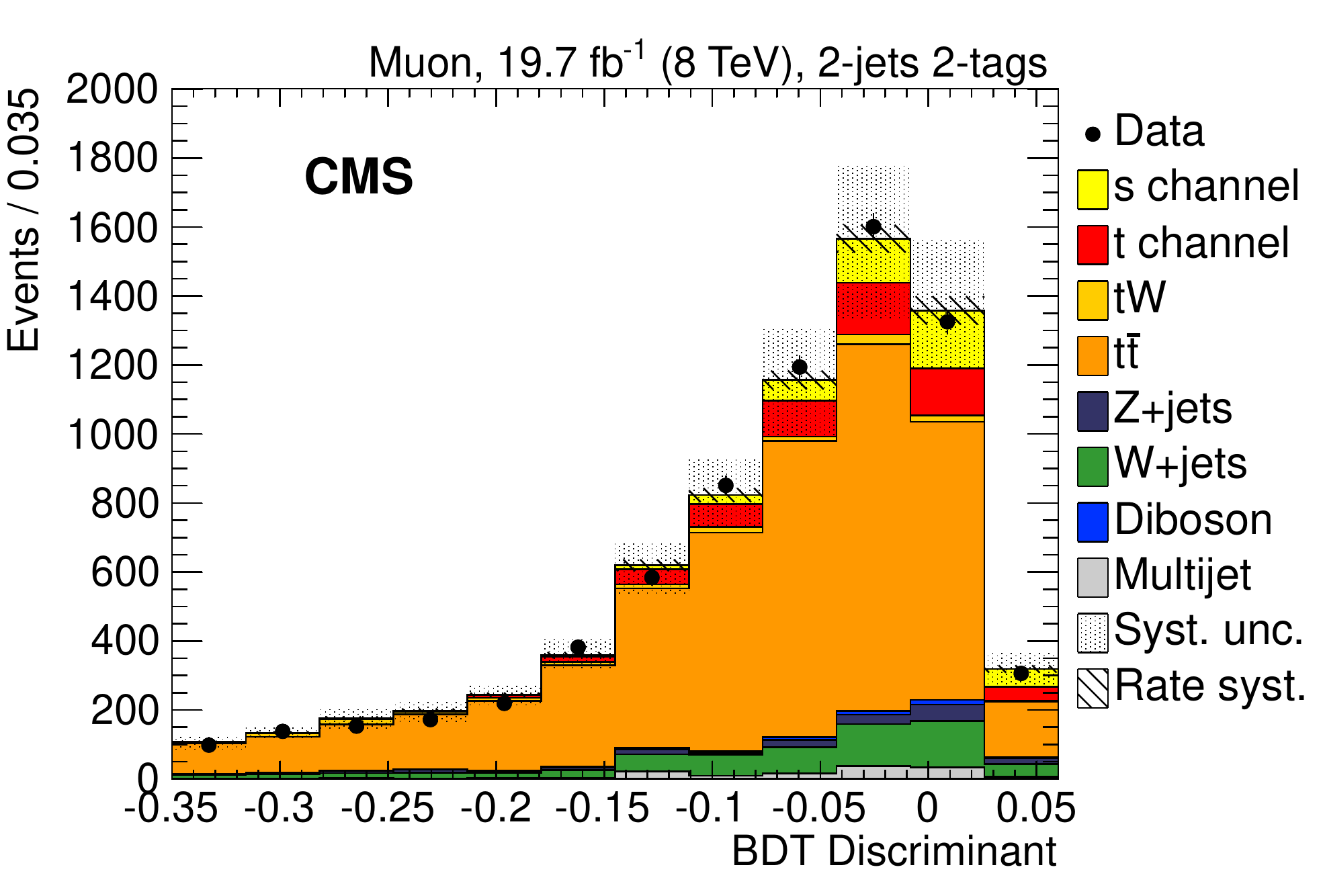}
           \includegraphics[width=0.48\textwidth]{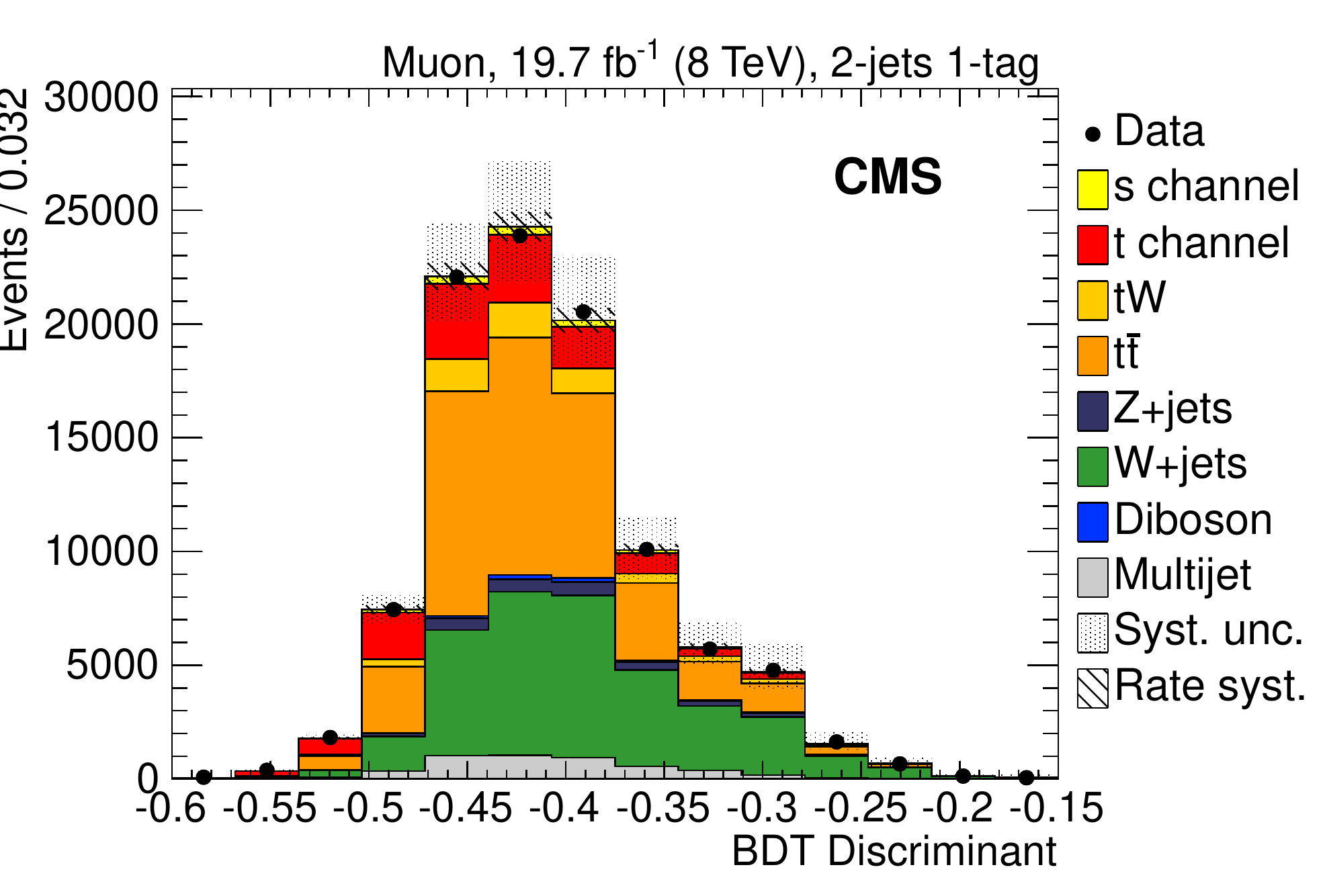}
           \includegraphics[width=0.48\textwidth]{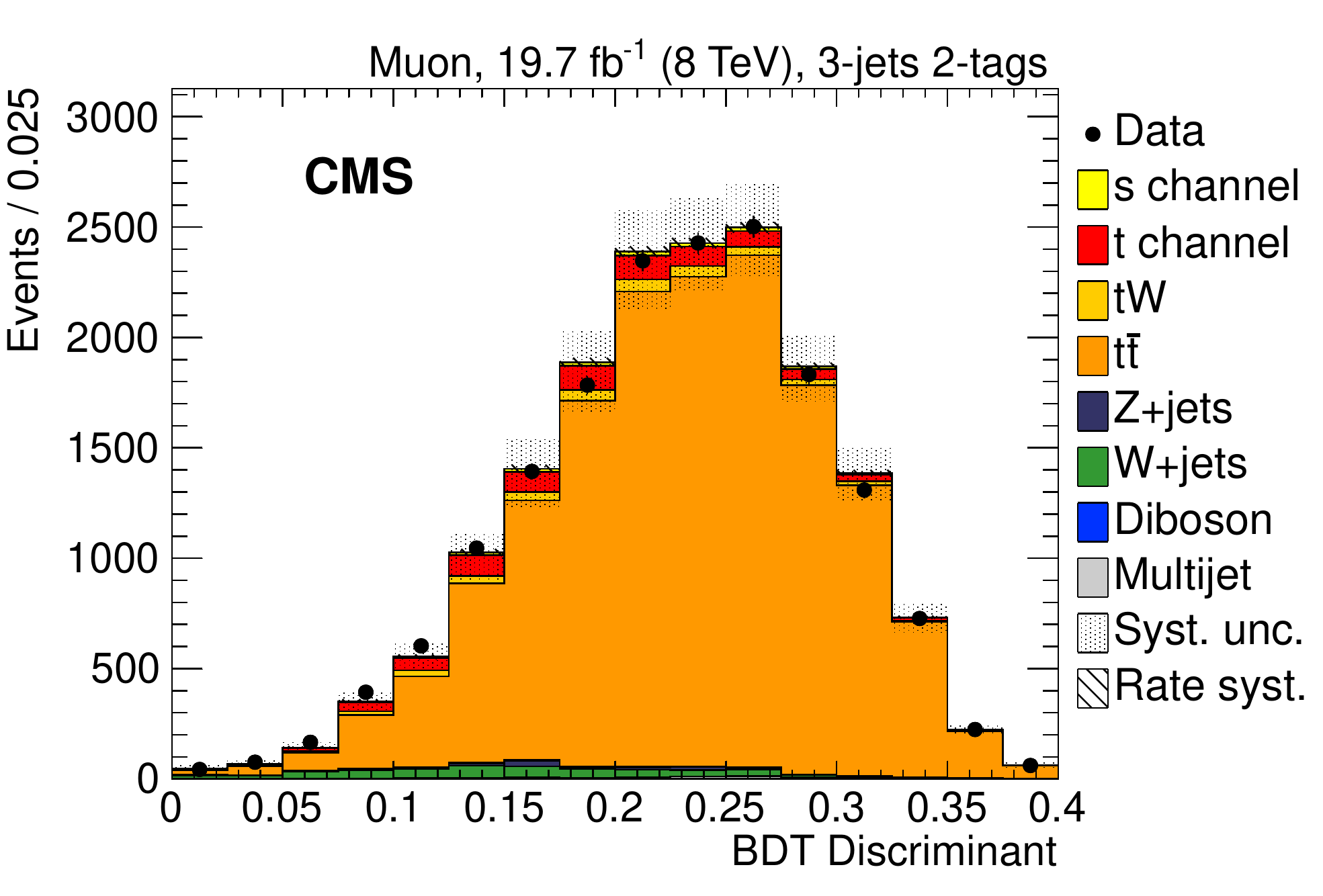}
            \caption{\label{fig:bdtdataMCMufit}Comparison of data with simulation for the distributions of the
              BDT discriminants in the (upper left) 2-jets 2-tags, (upper right) 2-jets 1-tag, and (bottom) 3-jets 2-tags event category,
              for the muon channel at 8\TeV. The simulation is normalized to the combined (7+8\TeV) fit results.
	      The inner uncertainty bands include the post-fit background normalizations uncertainties only, the outer ones
              include the total systematic uncertainty, obtained summing in quadrature the individual contributions.}
       \end{figure}

\begin{figure}[!htb]
         \centering
           \includegraphics[width=0.48\textwidth]{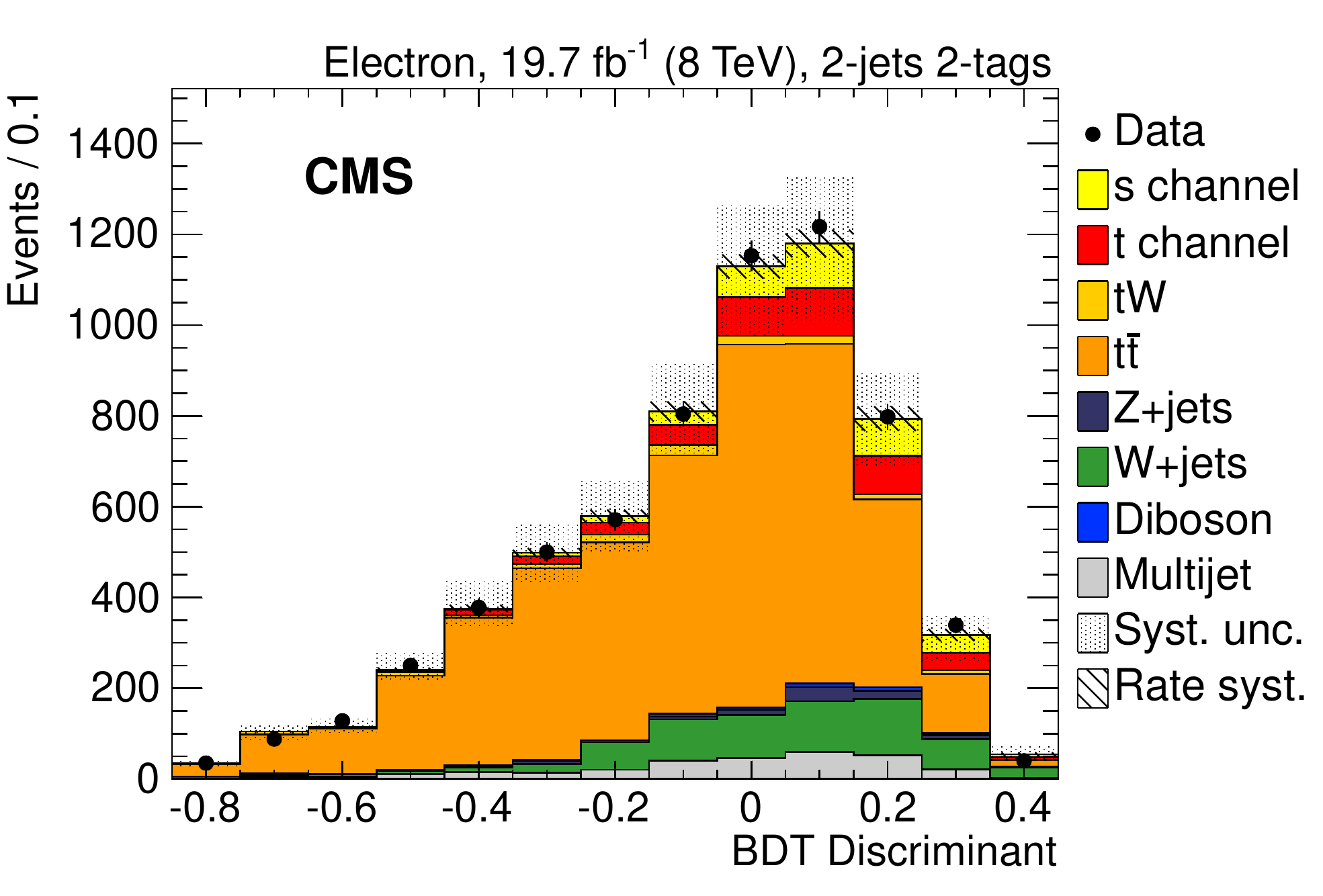}
           \includegraphics[width=0.48\textwidth]{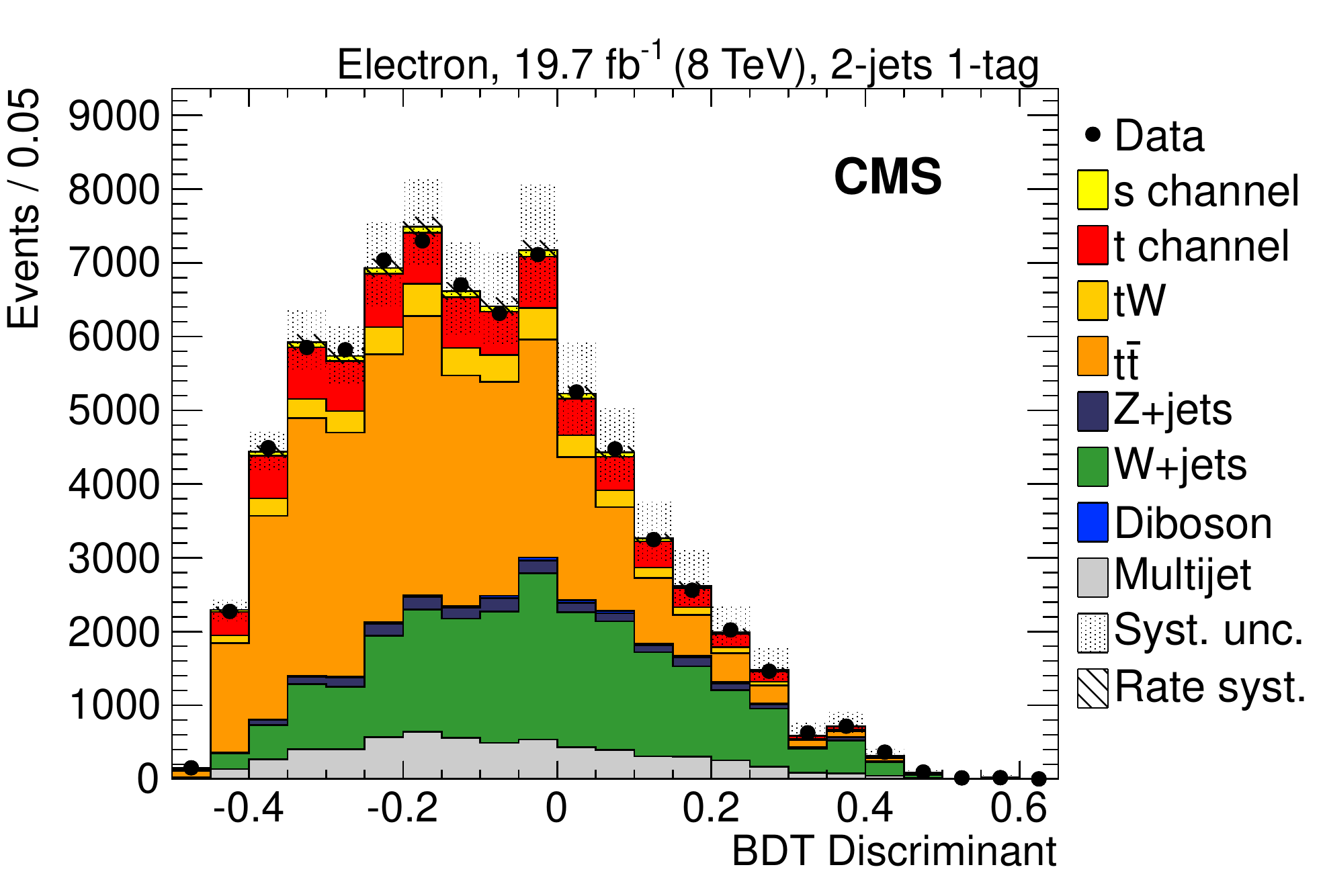}
           \includegraphics[width=0.48\textwidth]{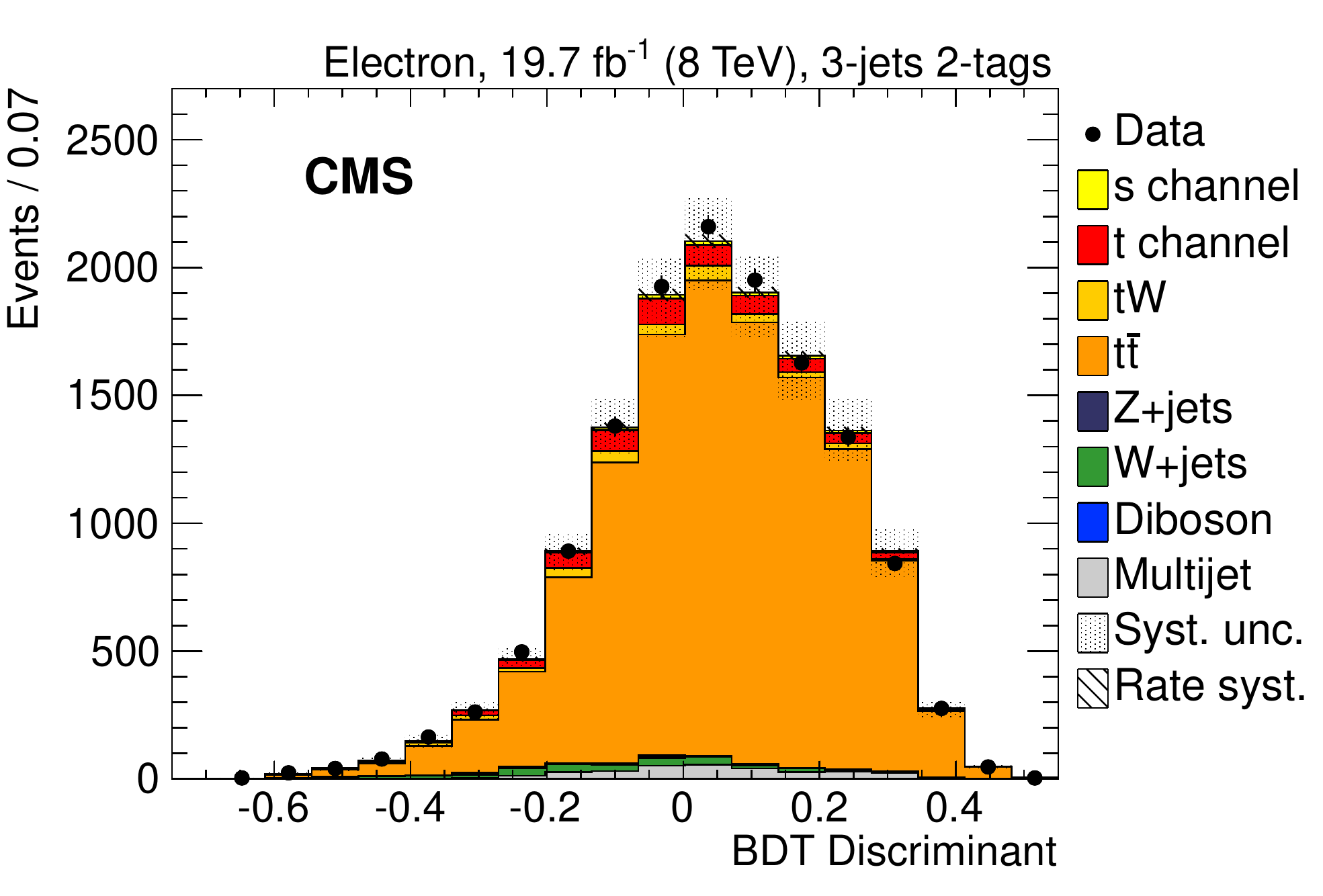}
           \caption{\label{fig:bdtdataMCElefit}Comparison of data with simulation for the distributions of the BDT discriminants
             in the (upper left) 2-jets 2-tags, (upper right) 2-jets 1-tag, and (bottom) 3-jets 2-tags event category, for the electron channel at 8\TeV.
             The simulation is normalized to the combined (7+8\TeV) fit results.
	     The inner uncertainty bands include the post-fit background normalizations uncertainties only,
             the outer ones include the total systematic uncertainty, obtained summing in quadrature the individual contributions.}
       \end{figure}

Figures~\ref{fig:bdtdataMCMufit7TeV},~\ref{fig:bdtdataMCMufit}, and~\ref{fig:bdtdataMCElefit} show the comparison of
the BDT discriminant distributions for all the event categories
in the muon channel at 7\TeV and muon and electron channels at 8\TeV, after the fit to the combined channels.
Tables~\ref{tab:7_8yield_2j2t},~\ref{tab:7_8yield_2j1t} and~\ref{tab:7_8yield_3j2t} 
summarize the number of events selected according to the requirements
described in Section~\ref{sec:selection}, including the requirement $\mTW>50$\GeV at 7\TeV 
in the 2-jets 1-tag category, and after the fit to the combined channels.
The SM expectation for the $s$ channel in the 2-jets 2-tags category
is 64 events selected in the muon channel at 7\TeV, 223 in the muon channel at 8\TeV, 
and 171 in the electron channel at 8\TeV.

\begin{table}
\centering
\topcaption{Event yields for the main processes in the 2-jets 2-tags region, at 7 and 8\TeV.
The yields of the simulated samples are quoted after the likelihood-maximization procedure 
for the combined 7+8\TeV fit. The uncertainties include the statistical uncertainty on the simulation, 
the background normalizations uncertainties and the b tagging uncertainty.
\label{tab:7_8yield_2j2t}}
\newcolumntype{x}{D{,}{\,\pm\,}{4.3}}
\begin{tabular}{ lxxx }
\hline
Process     & \multicolumn{1}{c}{$\mu$, 7\TeV}   & \multicolumn{1}{c}{$\mu$, 8\TeV}    & \multicolumn{1}{c}{e, 8\TeV}    \\
\hline
\ttbar      & 1380,80   & 4960,340   & 4290,300 \\
\wjets       & 150,30   & 580,110   & 620,110 \\
\zjets       & 22,7    & 160,40    & 90,30  \\
Diboson      & 3,3    & 59,16    & 46,13  \\
Multijet     & 70,20   & 130,40    & 290,60  \\
tW           & 37,6    & 149,19    & 130,16  \\
$t$ channel  & 135,16   & 570,50    & 420,40  \\[\cmsSkipLength]
$s$ channel  & 129,5    & 452,16   & 347,12   \\[\cmsSkipLength]
Total MC     & 1920,110  & 7060,370   & 6240,320  \\[\cmsSkipLength]
Data         & \multicolumn{1}{c}{1883}            & \multicolumn{1}{c}{7023}             & \multicolumn{1}{c}{6301}            \\
\hline
 \end{tabular}
 \end{table}

 \begin{table} 
 \centering
\topcaption{Event yields for the main processes in the 2-jets 1-tag region, at 7 and 8\TeV.
The yields of the simulated samples are quoted after the likelihood-maximization procedure 
for the combined 7+8~\TeV fit. The uncertainties include the statistical uncertainty on the simulation, 
the background normalizations uncertainties and the b tagging uncertainty. 
\label{tab:7_8yield_2j1t}}
\newcolumntype{x}{D{,}{\,\pm\,}{4.5}}
\begin{tabular}{ lxxx } 
\hline 
Process     & \multicolumn{1}{c}{$\mu$, 7\TeV}   & \multicolumn{1}{c}{$\mu$, 8\TeV}    & \multicolumn{1}{c}{e, 8\TeV}    \\
\hline 
\ttbar       & 6390,310   & 38900,1800    & 33200,910   \\ 
\wjets       & 4850,310   & 32900,1500    & 20090,940   \\
\zjets       & 240,50     & 2640,580      & 1820,390    \\    
Diboson      & 26,10      & 650,140       & 330,70      \\    
Multijet     & 78,78      & 4640,460      & 6080,300    \\   
tW           & 750,60     & 5380,460      & 3820,330    \\   
$t$ channel  & 2260,140   & 12730,760     & 7680,460    \\[\cmsSkipLength]
$s$ channel  & 281,5      & 1412,9        & 870,5       \\[\cmsSkipLength]
Total MC     & 14870,480  & 99240,2600    & 73900,1500  \\[\cmsSkipLength]
Data         & \multicolumn{1}{c}{14851}  & \multicolumn{1}{c}{99240}   & \multicolumn{1}{c}{73895}   \\
\hline 
 \end{tabular} 
 \end{table}

\begin{table}
\centering
\topcaption{Event yields for the main processes in the 3-jets 2-tags region, at 7 and 8\TeV.
The yields of the simulated samples are quoted after the likelihood-maximization procedure 
for the combined 7+8\TeV fit. The uncertainties include the statistical uncertainty on the simulation, 
the background normalizations uncertainties and the b tagging uncertainty.
\label{tab:7_8yield_3j2t}}
\newcolumntype{x}{D{,}{\,\pm\,}{4.4}}
\begin{tabular}{ lxxx }
\hline
Process     & \multicolumn{1}{c}{$\mu$, 7\TeV}   & \multicolumn{1}{c}{$\mu$, 8\TeV}    & \multicolumn{1}{c}{e, 8\TeV}    \\
\hline
\ttbar       & 3260,220   & 15200,900   & 12520,720 \\ 
\wjets       & 94,20      & 280,60      & 230,50    \\
\zjets       & 13,5       & 90,30       & 34,14     \\    
Diboson      &  0,0       & 24,6        & 17,4      \\    
Multijet     & 40,40      & 80,20       & 310,90    \\   
tW           & 78,13      & 370,60      & 320,50    \\   
$t$ channel  & 210,30     & 790,90      & 580,70    \\[\cmsSkipLength]
$s$ channel  & 38,2       & 126,5       & 94,4      \\[\cmsSkipLength]
Total MC     & 3730,230   & 16940,910   & 14120,730 \\[\cmsSkipLength]
Data         & \multicolumn{1}{c}{3848}  & \multicolumn{1}{c}{16934}   & \multicolumn{1}{c}{13512}   \\
\hline 
 \end{tabular}
 \end{table}

The sensitivity to the $s$ channel single top quark signal is estimated using the derivative of the likelihood test statistic,
defined as
\begin{equation*}
q_0 = \frac{\partial \text{log} L }{\partial \beta_{\text{signal}}} \bigg|_{\beta_{\text{signal}}=0},
\end{equation*}
and evaluated at the maximum-likelihood estimate in the background-only hypothesis.
Pseudo-data are generated to construct the distribution of the test statistic for the
background-only and the signal + background hypotheses. All the nuisance parameters are allowed
to vary according to their prior distributions in the pseudo-experiments,
while in the evaluation of $q_0$, the likelihood is maximized only with respect to
the background normalizations nuisance parameters.

\section{Results}
\label{sec:results}
The single top quark production cross section in the $s$ channel has been measured to be:
\begin{equation*}\begin{aligned}
\sigma_{s} &= \xsecmlemuseven \pm \xsecmlemusevenerrpb \,\text{(stat + syst)}\unit{pb,}   &&\text{muon channel, 7\TeV;} \\
\sigma_{s} &= \xsecmlemu \pm \xsecmlemuerrpb \,\text{(stat + syst)}\unit{pb,}     &&\text{muon channel, 8\TeV;} \\
\sigma_{s} &= \xsecmleele \pm \xsecmleeleerrpb \,\text{(stat + syst)}\unit{pb,}   &&\text{electron channel, 8\TeV;} \\
\sigma_{s} &= \xsecmlecomb \pm \xsecmlecomberrpb \,\text{(stat + syst)}\unit{pb,} &&\text{combined, 8\TeV.}
\end{aligned}\end{equation*}
The observed (expected) significance of the measurement is \sigobsseven(\sigexpseven) standard deviations at 7\TeV and \sigobscomb(\sigexpcomb)
for the combined muon and electron fit at 8\TeV. The 68\% CL interval for the expected significance is 0.0--1.5 at 7\TeV and
0.0--1.8 at 8\TeV.

The combined fit to the 7 and 8\TeV data determines the signal cross section relative to the SM predictions
with a best fit value of $\beta_{\text{signal}} = \xsecmlecombfinalSF \pm \xsecmlecombfinalSFunc$.
The observed significance of the measurement is \sigobscombfinal standard deviations with \sigexpcombfinal standard deviations expected.

The observed upper limit on the $s$ channel cross section at 95\% CL is \clsobsmusevenpb\unit{pb}at 7\TeV and
\clsobscombpb\unit{pb}for the combined muon and electron channel at 8\TeV. Combining the 7 and 8\TeV data, the observed upper limit on the signal strength is \clsobscombfinalSF.
In Table~\ref{tab:upperlimits}, we report a summary of the observed and expected upper limits
at 7 and 8\TeV and for the combination of the channels.

\begin{table}[!h]
\centering
\topcaption{Observed and expected upper limits (UL) at 7 and 8\TeV and for the combination of the data. Both the expected limits assuming the presence of a SM signal or in the absence of a signal are reported. In the hypothesis of a SM signal, the 68\% CL interval for the expected limit is also reported within square brackets. In the last row the upper limits are given in terms of the rate relative to the SM expectation.\label{tab:upperlimits}}
   \begin{tabular}{ lccc }
     \hline
     Channel & Observed UL & Expected UL---SM signal & Expected UL---no signal  \\
     \hline
      $\mu$, 7\TeV            & \clsobsmusevenpb\unit{pb}  & \clsexpmusevenpbsig [\clsexpmusevenpbsiginf, \clsexpmusevenpbsigsup]\unit{pb}     & \clsexpmusevenpbnosig\unit{pb}  \\
      $\mu+\Pe$, 8\TeV        & \clsobscombpb\unit{pb}  & \clsexpcombpbsig [\clsexpcombpbsiginf, \clsexpcombpbsigsup]\unit{pb}     & \clsexpcombpbnosig\unit{pb}  \\
      7$+$8\TeV              & \clsobscombfinalSF  & \clsexpcombfinalSFsig [\clsexpcombfinalSFsiginf, \clsexpcombfinalSFsigsup] & \clsexpcombfinalSFnosig  \\
\hline
\end{tabular}
\end{table}

\section{Summary}
\label{sec:conclusions}
A search is presented for single top quark production in the $s$ channel
in pp collisions at centre-of-mass energies of 7 and 8\TeV
with the CMS detector at the LHC.
A multivariate approach based on boosted decision trees is adopted
to discriminate the signal from background contributions.
The cross section is measured to be
$\xsecmlemuseven \pm \xsecmlemusevenerrpb$\,(stat + syst)\unit{pb}at 7\TeV and
$\xsecmlecomb \pm \xsecmlecomberrpb$\,(stat + syst)\unit{pb}at 8\TeV,
corresponding to a combined signal rate relative to SM expectations of
$\xsecmlecombfinalSF \pm \xsecmlecombfinalSFunc$\,(stat + syst).
The observed significance of the combined measurement is \sigobscombfinal standard deviations with \sigexpcombfinal
standard deviations expected. 
The observed and expected upper limits on the combined signal strength are found to be
\clsobscombfinalSF and \clsexpcombfinalSFsig at 95\% CL, respectively.
The measurements are in agreement with the prediction of the standard model.

\begin{acknowledgments}
We congratulate our colleagues in the CERN accelerator departments for the excellent performance of the LHC and thank the technical and administrative staffs at CERN and at other CMS institutes for their contributions to the success of the CMS effort. In addition, we gratefully acknowledge the computing centres and personnel of the Worldwide LHC Computing Grid for delivering so effectively the computing infrastructure essential to our analyses. Finally, we acknowledge the enduring support for the construction and operation of the LHC and the CMS detector provided by the following funding agencies: BMWFW and FWF (Austria); FNRS and FWO (Belgium); CNPq, CAPES, FAPERJ, and FAPESP (Brazil); MES (Bulgaria); CERN; CAS, MoST, and NSFC (China); COLCIENCIAS (Colombia); MSES and CSF (Croatia); RPF (Cyprus); MoER, ERC IUT and ERDF (Estonia); Academy of Finland, MEC, and HIP (Finland); CEA and CNRS/IN2P3 (France); BMBF, DFG, and HGF (Germany); GSRT (Greece); OTKA and NIH (Hungary); DAE and DST (India); IPM (Iran); SFI (Ireland); INFN (Italy); MSIP and NRF (Republic of Korea); LAS (Lithuania); MOE and UM (Malaysia); CINVESTAV, CONACYT, SEP, and UASLP-FAI (Mexico); MBIE (New Zealand); PAEC (Pakistan); MSHE and NSC (Poland); FCT (Portugal); JINR (Dubna); MON, RosAtom, RAS and RFBR (Russia); MESTD (Serbia); SEIDI and CPAN (Spain); Swiss Funding Agencies (Switzerland); MST (Taipei); ThEPCenter, IPST, STAR and NSTDA (Thailand); TUBITAK and TAEK (Turkey); NASU and SFFR (Ukraine); STFC (United Kingdom); DOE and NSF (USA).

Individuals have received support from the Marie-Curie programme and the European Research Council and EPLANET (European Union); the Leventis Foundation; the A. P. Sloan Foundation; the Alexander von Humboldt Foundation; the Belgian Federal Science Policy Office; the Fonds pour la Formation \`a la Recherche dans l'Industrie et dans l'Agriculture (FRIA-Belgium); the Agentschap voor Innovatie door Wetenschap en Technologie (IWT-Belgium); the Ministry of Education, Youth and Sports (MEYS) of the Czech Republic; the Council of Science and Industrial Research, India; the HOMING PLUS programme of the Foundation for Polish Science, cofinanced from European Union, Regional Development Fund; the OPUS programme of the National Science Center (Poland); the Compagnia di San Paolo (Torino); the Consorzio per la Fisica (Trieste); MIUR project 20108T4XTM (Italy); the Thalis and Aristeia programmes cofinanced by EU-ESF and the Greek NSRF; the National Priorities Research Program by Qatar National Research Fund; the Rachadapisek Sompot Fund for Postdoctoral Fellowship, Chulalongkorn University (Thailand); and the Welch Foundation, contract C-1845.
\end{acknowledgments}

\bibliography{auto_generated}

\cleardoublepage \appendix\section{The CMS Collaboration \label{app:collab}}\begin{sloppypar}\hyphenpenalty=5000\widowpenalty=500\clubpenalty=5000\textbf{Yerevan Physics Institute,  Yerevan,  Armenia}\\*[0pt]
V.~Khachatryan, A.M.~Sirunyan, A.~Tumasyan
\vskip\cmsinstskip
\textbf{Institut f\"{u}r Hochenergiephysik der OeAW,  Wien,  Austria}\\*[0pt]
W.~Adam, E.~Asilar, T.~Bergauer, J.~Brandstetter, E.~Brondolin, M.~Dragicevic, J.~Er\"{o}, M.~Flechl, M.~Friedl, R.~Fr\"{u}hwirth\cmsAuthorMark{1}, V.M.~Ghete, C.~Hartl, N.~H\"{o}rmann, J.~Hrubec, M.~Jeitler\cmsAuthorMark{1}, V.~Kn\"{u}nz, A.~K\"{o}nig, M.~Krammer\cmsAuthorMark{1}, I.~Kr\"{a}tschmer, D.~Liko, T.~Matsushita, I.~Mikulec, D.~Rabady\cmsAuthorMark{2}, N.~Rad, B.~Rahbaran, H.~Rohringer, J.~Schieck\cmsAuthorMark{1}, R.~Sch\"{o}fbeck, J.~Strauss, W.~Treberer-Treberspurg, W.~Waltenberger, C.-E.~Wulz\cmsAuthorMark{1}
% ...
\vskip\cmsinstskip
% ...
1:~~Also at Vienna University of Technology, Vienna, Austria\\
2:~~Also at CERN, European Organization for Nuclear Research, Geneva, Switzerland\\
% ...

\end{sloppypar}
\end{document}